\documentclass[twocolumn,showpacs,preprintnumbers,amsmath]{revtex4}
\usepackage{graphicx}
\usepackage{amsmath}

\begin{document}

\title{\textbf{Decoherence induced by electron accumulation \\
in quantum measurement of charge qubits}}
\author{\textbf{Ming-Tsung Lee$^{a}$ and Wei-Min Zhang$^{a,b}$}}
\date{\today}
\affiliation{$^{a}$Physics Division, National Center for Theoretical
Sciences,
Tainan, Taiwan 70101, ROC \\
$^{b}$Department of Physics and Center for Quantum Information
Science, National Cheng Kung University, Tainan, Taiwan 70101, ROC}

\begin{abstract}
In this paper, we study the quantum decoherence induced by
accumulation of electron tunnellings during the quantum measurement
of a charge qubit. The charge qubit is a single electron confined in
coupled quantum dots. The measurement of the qubit states is
performed using a quantum point contact. A set of master equations
for qubit states is derived within a non-equilibrium perturbation to
the equilibrium reservoir due to the electron accumulation between
the source and drain of the quantum point contact. The quantum
decoherence of the qubit states arose from the electron accumulation
during the measurement is explored in this framework, and several
interesting results on charge qubit decoherence are obtained.
\end{abstract}

\pacs{03.65.Yz,85.35.Be,03.65.Ta,03.67.Lx,73.63.Kv} \maketitle

\section{INTRODUCTION}

Quantum decoherence \cite{decoh1,decoh2,decoh3} is mainly induced by the
interaction of a microscopic system coupled with its environment. The
system-environment coupling results in a non-unitary evolution of the
system, which destroys the purity of quantum states and leads to information
loss toward the environment. These issues have attracted much attention in
the study of quantum computation and quantum communication in recent years
\cite{qis}. Some investigations of quantum decoherence have been focused on
the measurement induced decoherence \cite
{gurvitz,korotkov,goan,mozyrsky,clerk,stace,li}, and the dynamical controls
of the decoherence \cite{free decoh, dy decoup}. For the implementation of
realistic quantum information processors, these investigations become the
most champion works in the field. In the solid-state quantum computer with
charge qubits, quantum measurement of the qubit states can be performed by
coupling the charge qubit with a sensitive electrometer such as quantum
point contacts (QPCs) \cite{qpc} or a single-electron transistor \cite
{set,cooper}. In this paper, we shall focus on the charge qubit measurement
using a QPC, and study the non-equilibrium dynamical effect of the QPC on
the decoherence of the qubit states, here the charge qubit is a single
electron confined in coupled quantum dots (CQD).

Literaturely, to study the quantum decoherence induced by quantum
measurements, an equilibrium approximation is applied to the reservoir state
\cite{goan,stace,li}. For example, in the study of a charge qubit measured
by a QPC (Fig.~\ref{fig1}), the electronic reservoir, the source and the
drain of the QPC, is assumed to be macroscopic enough in comparison with the
CQD. Then the thermal equilibrium of the reservoir is kept continuously
through rapid relaxation processes. This condition of a perfect heat bath
causes electrons tunneling such that the extra electrons arriving at the
drain will flow back rapidly into the source through a close loop of the
transport circuit. No extra electron accumulates in the steady reservoir.
Practically, however, the condition of the perfect heat bath is not
guaranteed at mesoscopic scale of the QPC. The electron accumulation (EA) in
the source and the drain of the QPC may destroy the equilibrium of reservoir
and influence the outcome of the electrometer. As a result, the effects of
EA may become significant to the decoherence of the measured qubit. In this
paper, we shall treat approximately this intermediate non-equilibrium effect
as a perturbation to the equilibrium reservoir. The quantum decoherence of
the charge qubit due to the EA is then studied within this framework.

The paper is organized as follows: The model of the charge qubit measurement
and the perturbation scheme of taking the EA effect into account are
presented in Sec. II. A set of master equations for the reduced density
operator of the qubit is also derived in this section. In Sec. III., the
quantum decoherence of the qubit in the thermal equilibrium limit is
illuminated according to the resultant master equations. An abnormal
dependence of the qubit decay mode to the temperature and bias voltage is
obtained. The EA effect on the quantum decoherence of the qubit states is
explored in Sec. IV. The dynamics of the EA is analyzed, and the EA induced
decoherence of the qubit is extensively discussed. Remarkably, we also found
a decoherenceless EA effect in the low temperature regime. Finally, a
summary is given in Sec. V.

\section{CHARGE QUBIT MEASUREMENT USING QPC}

\begin{figure}[tbph]
\includegraphics*[angle=0,scale=.65]{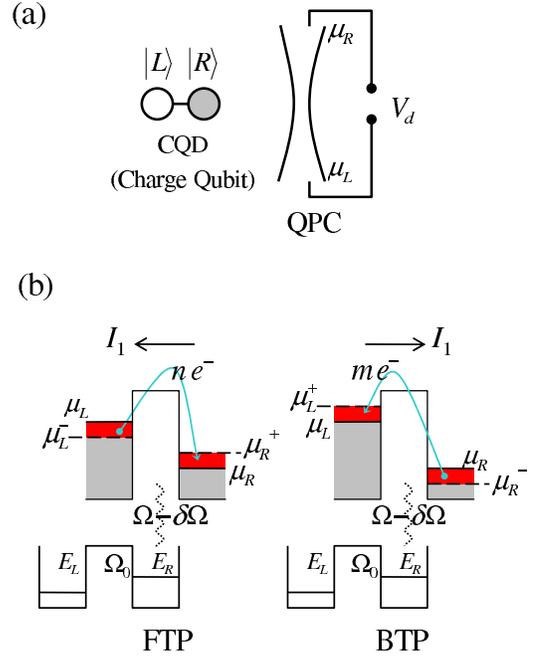}
\caption{(color online). (a) Schematic plot of a single electron in a
coupled quantum dots measured by the QPC. (b) Schematic diagram of the
electron accumulation in reservoir, the first figure shows the forward
tunneling processes (FTP) of electrons from the source to the drain which
causes electrons accumulating in the drain, and induces a variation of the
Fermi energy. The second one shows the backward tunneling processes (BTP) of
electrons from the drain to the source and a corresponding variation of the
Fermi energy.}
\label{fig1}
\end{figure}

The model we consider here is a single electron confined in a CQD (as a
charge qubit) that is measured by the QPC detector, see Fig.~\ref{fig1}a.
The Hamiltonian of the whole system is given by \cite
{gurvitz,korotkov,goan,stace,li}
\begin{eqnarray}
\hat{H} &=&\hat{H}_{S}+\hat{H}_{B}+\hat{H}^{\prime },  \label{eq:h1} \\
\hat{H}_{S} &=&E_{L}\hat{c}_{L}^{+}\hat{c}_{L}+E_{R}\hat{c}_{R}^{+}\hat{c}%
_{R}+\Omega _{0}\left( \hat{c}_{L}^{+}\hat{c}_{R}+\hat{c}_{R}^{+}\hat{c}%
_{L}\right) ,  \label{eq:hd} \\
\hat{H}_{B} &=&\sum_{l}\varepsilon _{l}\hat{a}_{l}^{+}\hat{a}%
_{l}+\sum_{r}\varepsilon _{r}\hat{a}_{r}^{+}\hat{a}_{r},  \label{eq:hpc} \\
\hat{H}^{\prime } &=&\left( \Omega -\delta \Omega \hat{c}_{R}^{+}\hat{c}%
_{R}\right) \sum_{lr}\left( \hat{a}_{l}^{+}\hat{a}_{r}+\hat{a}_{r}^{+}\hat{a}%
_{l}\right) .  \label{eq:hi}
\end{eqnarray}
Here, $\hat{H}_{S}$ is the Hamiltonian of the charge qubit in which $\hat{c}%
_{L}^{+}\left( \hat{c}_{L}\right) $ and $\hat{c}_{R}^{+}\left( \hat{c}%
_{R}\right) $ are the electron creation (annihilation) operators of the
electron sited in the two dots labelled by $L$ and $R$ as shown in Fig. 1a.
They satisfy the condition $\hat{c}_{L}^{+}\hat{c}_{L}+\hat{c}_{R}^{+}\hat{c}%
_{R}=1$. The parameter $\Omega _{0}$ is the electron tunneling amplitude
between the two dots. $\hat{H}_{B}$ denotes the Hamiltonian of the QPC
reservoir, which is decomposed into two terms for the source and the drain,
respectively, and $\hat{a}_{l}^{+}\left( \hat{a}_{l}\right) $ and $\hat{a}%
_{r}^{+}\left( \hat{a}_{r}\right) $ are the corresponding electron creation
(annihilation) operators. Since the presence of an electron in a dot close
proximity to the QPC will cause a variation in barrier of the QPC, $\Omega
\rightarrow \Omega -\delta \Omega $, the interaction Hamiltonian $\hat{H}%
^{\prime }$ describes the coupling of the quantum dots and the QPC in close
proximity to the dot $R$. The information of the qubit-states can then be
read out through the QPC current \cite{gurvitz}.

There are two channels for electron tunneling in the QPC. One is the forward
tunneling processes from the source to the drain, and the other is the
backward tunneling processes from the drain to the source, see Fig.~\ref
{fig1}b. In the high-bias regime, magnitude of the electron tunneling rate
for the forward tunneling processes is much larger than the one in the
backward tunneling processes. The contribution of the backward tunneling
processes could be ignored as treated in \cite{gurvitz}. However, in a
non-equilibrium description, both contributions of the forward tunneling
processes and backward tunneling processes should be included for an
arbitrary bias voltage. The Hilbert space is then spanned by the following
basis
\begin{equation}
\left\{ \left| \phi _{i}\right\rangle \otimes \left| D\left( \bar{L}%
^{n},R^{n}\right) \right\rangle ,\left| \phi _{i}\right\rangle \otimes
\left| \bar{D}\left( L^{m},\bar{R}^{m}\right) \right\rangle \right\} ,
\label{eq:basis}
\end{equation}
where $|\phi _{i}\rangle $ is an electron eigen-state in the CQD (we define $%
|\phi _{0}\rangle =|g\rangle $ and $|\phi _{1}\rangle =|e\rangle $ as the
ground and excited states in the eigen-state representation, respectively), $%
|D(\bar{L}^{n},R^{n})\rangle $$=\hat{a}_{r_{1}}^{+}\cdots \hat{a}_{r_{n}}^{+}%
\hat{a}_{\bar{l}_{1}}\cdots \hat{a}_{\bar{l}_{n}}\left| B_{0}\right\rangle $
is a state with $n$ electrons accumulated in the drain through the forward
tunneling processes, and $\left| \bar{D}\left( L^{m},\bar{R}^{m}\right)
\right\rangle =\hat{a}_{l_{1}}^{+}\cdots \hat{a}_{l_{m}}^{+}\hat{a}_{\bar{r}%
_{1}}\cdots \hat{a}_{\bar{r}_{m}}\left| B_{0}\right\rangle $ a state with $m$
electrons accumulated in the source through the backward tunneling
processes. The state $|B_{0}\rangle $ is the reservoir (equilibrium) vacuum,
and the operator $\hat{a}_{\bar{l}(\bar{r})}$ denotes the annihilation of an
electron from $|B_{0}\rangle $ [or the creation of a hole below the Fermi
energy of the source (drain)], while the operator $\hat{a}_{l(r)}^{+}$
denotes the creation of an electron above the Fermi energy of the source
(drain). Therefore, an arbitrary state of the whole system can be expressed
as
\begin{eqnarray}
|\psi (t)\rangle &=&\sum_{i}|\phi _{i}\rangle \otimes \Big(\sum_{n}\sum_{%
\bar{L}^{n}R^{n}}b_{i,\bar{L}^{n}R^{n}}(t)|D(\bar{L}^{n},R^{n})\rangle
\nonumber \\
&&+\sum_{m}\sum_{L^{m}\bar{R}^{m}}\tilde{b}_{i,L^{m}\bar{R}^{m}}(t)|\bar{D}%
(L^{m},\bar{R}^{m})\rangle \Big).  \label{eq:tstat}
\end{eqnarray}

To study the charge qubit measured by the QPC, the electronic reservoir is
usually assumed to be in the reservoir vacuum state (a thermal equilibrium
state). The reservoir vacuum state is uniquely determined for a given
temperature and chemical potential. The reservoir time correlation function
is usually calculated in this state. However, tunneling electrons from the
forward tunneling processes will occupy the levels above the Fermi energy of
the drain for a short period and disturbs the reservoir vacuum state, see
Fig.~\ref{fig1}b. This phenomenon, called the electron accumulation (EA) in
QPC, should be taken into account, vice versa for the backward tunneling
processes. Meanwhile, the EA may also be induced by the impurity of the
reservoir or a circuit with a worse transport mobility (the imperfect
condition of the reservoir). Literaturely, the effect of the EA is ignored
\cite{gurvitz,goan,mozyrsky,stace,li}. One assumes that electrons created in
drain (for the forward tunneling processes) will be forced into the circuit
instantly, no electrons are accumulated in the drain temporarily. Therefore,
there is no any variation to the Fermi energy. The read-out current is
completely contributed by the created electron of drain through the circuit,
as a result in the macroscopic limit. However, for a mesoscopic reservoir, a
variation of the Fermi energy can be generated by the EA. It is certainly
interesting to study what the influence of the EA effect is on the
decoherence of the measured qubit.

In general, the EA should be considered by treating the reservoir as a fully
non-equilibrium system. However, the reservoir is indeed an asymptotic
stationary state at mesoscopic scale. This is because the external bias can
be regarded as a bigger macroscopic reservoir compared with the QPC (see
Fig.~\ref{fig1}). The external bias forces the QPC reservoir to evolve into
an asymptotic stationary state in a short period. For such a case, we can
treat approximately the intermediate non-equilibrium effect as a
perturbation to the equilibrium reservoir. Thus, the EA can be described
simply by a variation of the chemical potentials in the source and the
drain. In the forward tunneling processes, the tunneling of $n$ electrons
moved from the source to the drain results in an effective increase of the
chemical potential in the drain $\mu _{R}\rightarrow \mu _{R}^{+}=\mu
_{R}+\delta \mu _{R}\left( n,\beta \right) $ and an effective decrease of
the chemical potential in the source $\mu _{L}\rightarrow \mu _{L}^{-}=\mu
_{L}-\delta \mu _{L}\left( n,\beta \right) $, see Fig.~\ref{fig1}b. Here, $%
\beta =1/k_{B}T$ is the inverse temperature. In thermal equilibrium state,
the electron distributions obey Fermi-Dirac function $F_{l}=\frac{1}{1+\exp
\beta \left( \varepsilon _{l}-\mu _{L}\right) }$ for the source and $F_{r}=%
\frac{1}{1+\exp \beta \left( \varepsilon _{r}-\mu _{R}\right) }$ for the
drain. Obviously, at a given temperature, the chemical potential determines
the number density of electrons $\bar{N}$ in reservoir. The relation of its
inverse function can be solved. Furthermore, because the QPC is a
2-dimensional electron gas, the density of states, $g_{L}$, for the source ($%
g_{R}$ for the drain) can be assumed to be energy independent \cite{gurvitz}%
. Then the chemical potential can be expressed as $\mu _{L,R}(\bar{N},\beta
)=\frac{1}{\mathbf{\beta }}\ln \left[ \exp (\beta \bar{N}/g_{L,R})-1\right] $%
. The variation due to the EA is given by $\mu _{R,L}(\bar{N}\pm n/A,\beta
)-\mu _{R,L}(\bar{N},\beta )=\pm \delta \tilde{\mu}_{R,L}(n,\beta )$, where $%
A$ is the area of the QPC. One can find that up to the first order of $n/A$,
\begin{equation}
\delta \tilde{\mu}_{R,L}(n,\beta )=\frac{n}{A}\left( \frac{1+e^{-\beta \mu
_{R,L}}}{g_{R,L}}\right) .  \label{eq:variatn}
\end{equation}

In addition, the bias effect on the EA should be taken into account. The
external bias and the QPC-dots interaction determine electron tunnelings in
the QPC. The current partially induced by the QPC-dots interaction records
the qubit information. The bias effect simply forces the reservoir into a
local equilibrium state. That is, the fluctuation of the chemical potential
arisen from the QPC-dots interaction will be suppressed by the bias effect
asymptotically, $\mu _{L}\left( t\right) -\mu _{R}\left( t\right)
\rightarrow V_{d}$. The EA should decay to vanish (a electron-release
process, ER for short) by the bias effect. This ER process is also
associated with the imperfect condition of the reservoir. To simplify the
problem, the stochastic EA process (also including the effects of the bias
and QPC-dots interaction) is assumed to be described by
\begin{equation}
\delta \mu _{R,L}\left( n,\beta ,t\right) =\delta \tilde{\mu}_{R,L}\left(
n,\beta \right) e^{-\frac{\Gamma _{R,L}\left( n\right) t}{\hbar }}\cos
\left( \omega _{d}t\right) .  \label{eq:verls}
\end{equation}
The exponential decay factor in Eq. (\ref{eq:verls}) comes from the bias
effect. The detailed expression of the ER decay rate $\Gamma _{R,L}\left(
n\right) $ depends on the imperfect condition of the QPC associated with the
bias effect. In general, $\Gamma _{R,L}$ are functions of the tunneling
electron number $n$, and are different for the forward tunneling processes
and backward tunneling processes. $1/\Gamma _{R,L}\left( n\right) $ denotes
the release time of the ER. The processes repeat continually through the
transport circuit which is given by the oscillation function in Eq. (\ref
{eq:verls}), where $\omega _{d}$ is the frequency of the repeated processes.
It can be estimated that $\omega _{d}\sim v_{t}/l$ with $l$ the length of
the QPC in the tunneling direction and $v_{t}$ the electron drift velocity
in the transport circuit. Thus, Eq. (\ref{eq:verls}) describes properly the
stochastic nature of the tunneling processes embedded in the ER process and
the cyclic motion of the external transport circuit. Obviously, the
magnitude of the chemical potential variation decays to zero as the
reservoir approaches to the local equilibrium state, namely, $\mu _{L}\left(
t\right) -\mu _{R}\left( t\right) =V_{d}$. Similar to the forward tunneling
processes, the tunneling of $m$ electrons moved from the drain to the source
for the backward tunneling processes leads to the chemical potential
decreasing $\mu _{R}\rightarrow \mu _{R}^{-}=\mu _{R}-\delta \mu
_{R}(m,\beta ,t)$ in the drain and increasing $\mu _{L}\rightarrow \mu
_{L}^{+}=\mu _{L}+\delta \mu _{L}(m,\beta ,t)$ in the source. Up to the
first order of $\delta \mu _{R,L}\left( n,\beta ,t\right) $, the perturbed
Fermi-Dirac function is given by
\begin{equation}
F_{r,l}^{\pm }(n,\beta ,t)\approx F_{r,l}\pm \delta \mu _{R,L}(n,\beta ,t)%
\frac{\partial F_{r,l}}{\partial \mu _{R,L}}.  \label{eq:pf}
\end{equation}

As a result, even if the EA disturbs slightly the equilibrium of the
reservoir, it may induce decoherence to the measured qubit. This effect can
be explored from the master equation of the measured qubit which has the
form as
\begin{equation}
\frac{d}{dt}\hat{\rho}\left( t\right) =\frac{1}{i\hbar }\left[ \hat{H}_{S},%
\hat{\rho}\left( t\right) \right] -\hat{R}\hat{\rho}\left( t\right) ,
\label{eq:mr}
\end{equation}
where $\hat{\rho}(t)$ is the reduced density operator of the qubit
\begin{equation}
\hat{\rho}(t)=\sum_{n=0}^{\infty }\left( \mathrm{Tr}_{D^{(n)}}[\hat{\rho}%
_{tot}(t)]+\mathrm{Tr}_{\bar{D}^{(n)}}[\hat{\rho}_{tot}(t)]\right) .
\label{eq:dentot}
\end{equation}
$\hat{\rho}_{tot}(t)$ in Eq. (\ref{eq:dentot}) denotes the total density
operator of the whole system, and the conditional partial traces are defined
by $\mathrm{Tr}_{D^{(n)}}[\hat{O}]={\sum_{\bar{L}^{n},R^{n}}\langle D(\bar{L}%
^{n},R^{n})|\hat{O}|D(\bar{L}^{n},R^{n})\rangle }$ and $\mathrm{Tr}_{\bar{D}%
^{(m)}}[\hat{O}]={\sum_{L^{m},\bar{R}^{m}}\langle \bar{D}(L^{m},\bar{R}^{m})|%
\hat{O}|\bar{D}(L^{m},\bar{R}^{m})\rangle .}$ Here, $\mathrm{Tr}_{D^{\left(
n\right) }}$ ($\mathrm{Tr}_{\bar{D}^{(m)}}$) means to sum over all the
allowed energy levels of the reservoir with the condition of $n$ electrons
tunneling the QPC barrier through the forward (backward) tunneling processes
and then accumulating in the drain (source). $\hat{R}$ in Eq. (\ref{eq:mr})
denotes the dissipation operator, the detailed form of which can be found in
the Appendix. The qubit decoherence due to the electrical measurement is
governed by the dissipation term $-\hat{R}\hat{\rho}\left( t\right) $ in Eq. (\ref{eq:mr}%
). This term is composed of the reservoir spectrum functions, which are the
Fourier transformation of the reservoir time correlation functions defined
by \textrm{Tr}$_{D^{(n)}(\bar{D}^{(m)})}\big[\hat{f}^{+}(t)\hat{f}(t^{\prime
})\hat{\rho}_{tot}\big]$ and $\mathrm{Tr}_{D^{(n)}(\bar{D}^{(m)})}\big[\hat{f%
}(t)\hat{f}^{+}(t^{\prime })\hat{\rho}_{tot}\big]$. The reservoir
fluctuation due to the QPC-dots interaction is encoded in the reservoir
operators $\hat{f}^{+}(t)\hat{f}(t^{\prime })$ and $\hat{f}(t)\hat{f}%
^{+}(t^{\prime })$, where $\hat{f_{t}}={\sum_{lr}e^{i(\varepsilon
_{r}-\varepsilon _{l})t/\hbar }\hat{a}_{r}^{+}\hat{a}_{l}}$ is the electron
tunneling operator in the interaction picture of the reservoir. The
tunneling processes of electrons relative to the EA have also been taken
into account in these functions by treating the EA-fluctuated reservoir
state as the perturbed Fermi-Dirac function through Eq. (\ref{eq:pf}). As a
result, the decoherence of the measured qubit due to the EA effect is
manifested in the master equation through the first order perturbation of
the reservoir spectrum functions, which is associated with the perturbed
term $\pm \delta \mu _{R,L}(n,\beta ,t)\frac{\partial F_{r,l}}{\partial \mu
_{R,L}}$ of the perturbed Fermi-Dirac function in Eq. (\ref{eq:pf}). A
perturbation scheme taking the EA effect into account is then given as
follows

\begin{equation}
\hat{\rho}_{tot}(t)=\hat{\rho}_{0,tot}(t)+\xi \hat{\rho}_{1,tot}(t)+\xi ^{2}%
\hat{\rho}_{2,tot}(t)+\cdots ,  \label{eq:pertot}
\end{equation}
where $\xi =\mathcal{U}/\bar{\mu}$ is a perturbation parameter
characterizing the EA effect with $\mathcal{U}=\left( (\delta \tilde{\mu}%
_{L}(n,\beta )+\delta \tilde{\mu}_{R}(n,\beta )\right) /n$ and $\bar{\mu}%
=(\mu _{L}+\mu _{R})/2$. The zero order term $\hat{\rho}_{0,tot}(t)$ in Eq.~(%
\ref{eq:pertot}) represents the usual states without the EA. The higher
order terms describe the contribution of the EA. By taking partial trace to
Eq.~(\ref{eq:pertot}), we have the perturbation scheme for the reduced
density operator of the qubit
\begin{equation}
\hat{\rho}\left( t\right) =\hat{\rho}_{0}(t)+\xi \hat{\rho}_{1}(t)+\cdots .
\end{equation}

The QPC, as a sensitive charge detector, may be required a large electron
transparency and a relatively strong coupling strength to maximize the
detector sensitivity in the current experiments \cite{clerk,elzerman,averin}%
. However, we shall focus in this paper on the tunneling
transparency regime (the weak dot-QPC coupling). This is because,
as Gurvitz has pointed out, the occupation probability of a single
electron in the coupled dot (as a qubit) can be traced through the
QPC readout current in this regime \cite {gurvitz}. To take the
weak dot-QPC coupling into account, the second order cummulant
expansion technique \cite{mori,li,goan,stace} is used. Here, the
Markovian approximation has been introduced in the technique. It
should be noted that since the electron tunneling in the QPC is
determined by the external bias and the QPC-dots interaction, it
results in the qubit dynamics with three time scales: the qubit
decoherence time, the ER decay time due to the electron release by
the bias, and the correction time due to the QPC-dots interaction
associated with the bias, temperature and the electron
accumulation. In literature, the validity of the Markovian
approximation could depend on the time scale of the correlation
time \cite{gardiner}. When the correlation time is much smaller
than other time scales (in the Markovian regime), the
non-Markovian memory effect in the qubit dynamics can be ignored.
The qubit decoherence under the environment effect through the
QPC-dots interaction is simply a decay process. Taking the
electron accumulation effect into account, an extra qubit
relaxation may be induced but the enhancement should not be so
significant under the perturbative treatment. Therefore, the
Markovian approximation \cite{gardiner} is still applicable for
the derivation of the master equations here.

As a result, the master equations for the measured qubit in our perturbation
scheme are obtained as follows
\begin{eqnarray}
\xi ^{0} &:&\frac{d}{dt}\hat{\rho}_{0}(t)=-i\hat{L}_{D}\hat{\rho}%
_{0}(t)-\lambda \left[ \hat{q},\Big[\left[ \hat{G}^{(0)},\hat{\rho}%
_{0}(t)\right] \Big]\right] , \nonumber \\
\label{eq:ml0} \\
\xi ^{1} &:&\frac{d}{dt}\hat{\rho}_{1}(t)=-i\hat{L}_{D}\hat{\rho}%
_{1}(t)-\lambda \Big[\hat{q},\Big(\Big[\left[ \hat{G}^{(0)},\hat{\rho}%
_{1}(t)\right] \Big]  \nonumber \\
&&~~~~~~~~~~~~~~~~~~+\mathcal{F}\left( t\right) \Big[\left[ \hat{G}^{(1)},%
\hat{N}_{0}(t)\right] \Big]\Big)\Big].  \label{eq:ml1rel}
\end{eqnarray}
Eq. (\ref{eq:ml0}) is nothing but the rate equation of the qubit for the
reservoir in the thermal equilibrium state without the EA effect. The EA
effect on the measured qubit is described by Eq. (\ref{eq:ml1rel}). In Eqs. (%
\ref{eq:ml0},\ref{eq:ml1rel}), $\hat{L}_{D}$ is the Liouvillian operator for
the qubit, $\lambda =\pi g_{L}g_{R}/\hbar $, the double-bracket commutator $%
\left[ \left[ \hat{A},\hat{B}\right] \right] \equiv \hat{A}\hat{B}-\left(
\hat{A}\hat{B}\right) ^{\dagger }$, and $\mathcal{F}(t)$ represents the
contribution of the bias effect to the EA
\begin{equation}
\mathcal{F}\left( t\right) =\cos \left( \omega _{d}t\right) \frac{e^{-\frac{%
\Gamma _{R}t}{\hbar }}+re^{-\frac{\Gamma _{L}t}{\hbar }}}{1+r},r=\frac{%
g_{R}\left( 1+e^{-\beta \mu _{L}}\right) }{g_{L}\left( 1+e^{-\beta \mu
_{R}}\right) }.  \label{eq:replac3}
\end{equation}
Also, the operator $\hat{q}$ and $\hat{G}^{(k)}$ for $k=0,1$ in Eqs. (\ref
{eq:ml0},\ref{eq:ml1rel}) are defined as
\begin{eqnarray}
\hat{q} &=&\hat{P}_{0}-\hat{P}_{1}-\hat{P}_{2}, \nonumber  \\
\hat{G}^{(k)}
&=&\sum_{i=0,1,2}(G_{+,i}^{(k)}+G_{-,i}^{(k)})\hat{P}_{i},
\label{eq:qg}
\end{eqnarray}
and the operators $\hat{P}_{0,1,2}$ \cite{stace} are given by
\begin{eqnarray}
&&\hat{P}_{0}=\left( \Omega -\frac{\delta \Omega }{2}\right) 1+\frac{\delta
\Omega \cos \theta }{2}\left( \left| g\right\rangle \left\langle g\right|
-\left| e\right\rangle \left\langle e\right| \right) ,  \nonumber \\
&&\hat{P}_{1}=\frac{\delta \Omega \sin \theta }{2}\left| e\right\rangle
\left\langle g\right| ,\;\hat{P}_{2}=\frac{\delta \Omega \sin \theta }{2}%
\left| g\right\rangle \left\langle e\right| ,  \label{eq:p012}
\end{eqnarray}
where $\theta =\cos ^{-1}\left[ \left( E_{R}-E_{L}\right) /\gamma
\right] $, $E_{R,L}$ denote the energy for a single electron state
in right and left dot, and $\gamma =\sqrt{4\Omega _{0}^{2}+\left(
E_{R}-E_{L}\right) ^{2}}$ is the energy difference of the two
eigenstates of the charge qubit. The coupled dots structure is
shown by $\theta $. The operators $\hat{P}_{1,2}$ are responsible
for the inelastic excitation and relaxation of the electron in the
dots coupling with the QPC. The coefficients $G_{\pm ,i}^{\left(
0\right) }$ and $G_{\pm ,i}^{\left( 1\right) }$ in Eq.
(\ref{eq:qg})\ are the reservoir spectrum functions without and
with the EA
\begin{eqnarray}
&&G_{+,0}^{(0)}=\tilde{g}^{(0)}\left( eV_{d}\right) ,\;G_{+,0}^{(1)}=-\bar{%
\mu}\tilde{g}^{(1)}\left( eV_{d}\right) ,  \nonumber \\
&&G_{+,1}^{(0)}=-\tilde{g}^{(0)}\left( eV_{d}-\gamma \right)
,\;G_{+,1}^{(1)}=\bar{\mu}\tilde{g}^{(1)}\left( eV_{d}-\gamma \right) ,\;
\nonumber \\
&&G_{+,2}^{(0)}=-\tilde{g}^{(0)}\left( eV_{d}+\gamma \right) ,G_{+,2}^{(1)}=%
\bar{\mu}\tilde{g}^{(1)}\left( eV_{d}+\gamma \right) ,  \label{eq:bg1} \\
&&G_{-,i}^{(0)}=\left. G_{+,i}^{(0)}\right| _{V_{d}\rightarrow
-V_{d}},\;G_{-,i}^{(1)}=-\left. G_{+,i}^{(1)}\right| _{V_{d}\rightarrow
-V_{d}},  \label{eq:bgm}
\end{eqnarray}
and
\begin{equation}
\tilde{g}^{(0)}(x)=\frac{x}{1-e^{-\beta x}},\;\tilde{g}^{(1)}(x)=\frac{%
\partial }{\partial \beta }\left( \frac{\beta }{1-e^{-\beta x}}\right) ,
\label{eq:g01}
\end{equation}
where $V_{d}$ is the bias voltage and $e$ the electron charge. For
convenience, $e=1$ is used hereafter. Also, $\hat{N}_{0}(t)$ in Eq. (\ref
{eq:ml1rel}) is the number density operator of the tunneling electrons for
the reservoir in the thermal equilibrium state \cite{gurvitz}.

Meanwhile, the qubit information can be read out through the QPC\ current $%
\hat{I}=\frac{d\hat{N}}{dt}$, where $\hat{N}={\sum_{n=0}^{\infty }}n\mathrm{%
Tr}_{D^{(n)}}[\hat{\rho}_{tot}(t)]-{\sum_{m=0}^{\infty }}m\mathrm{Tr}_{\bar{D%
}^{(m)}}[\hat{\rho}_{tot}(t)].$ An extra minus sign in the second term of $%
\hat{N}$ is added because electrons in the backward tunneling processes
tunnel backwardly from drain to source. In our perturbation scheme, $\hat{I}=%
\frac{d}{dt}\hat{N}_{0}+\xi \frac{d}{dt}\hat{N}_{1}+\mathcal{O(}\xi
^{2})+\cdots $, and $\hat{N}_{1}(t)$ is the number density operator of the
tunneling electrons for the reservoir fluctuated by the EA. The QPC current
is governed by

\begin{eqnarray}
\xi ^{0} &:&\frac{d}{dt}\hat{N}_{0}(t)=-i\hat{L}_{D}\hat{N}_{0}(t)-\lambda %
\Big\{\left[ \hat{q},\left[ \left[ \hat{G}^{(0)},\hat{N}_{0}(t)\right]
\right] \right]   \nonumber \\
&&~~~~~~~~~~~~~~~~~~~~~~~-\left( \hat{\overline{G}}^{(0)}\hat{\rho}_{0}(t)%
\hat{q}+H.c.\right) \Big\},  \label{eq:mc0} \\
\xi ^{1} &:&\frac{d}{dt}\hat{N}_{1}(t)=-i\hat{L}_{D}\hat{N}_{1}(t)-\lambda %
\Big\{\Big[\hat{q},\Big(\left[ \left[ \hat{G}^{(0)},\hat{N}_{1}(t)\right]
\right]   \nonumber \\
&&~~~~~~~~~~~~~~~~~~~~~+\mathcal{F}\left( t\right) \left[ \left[ \hat{G}%
^{(1)},\hat{W}_{0}(t)\right] \right] \Big)\Big]  \nonumber \\
&&-\Big( \Big( \hat{\overline{G}}^{(0)}\hat{\rho}_{1}(t)+\mathcal{F}(t)\hat{%
\overline{G}}^{(1)}\hat{N}_{0}(t)\Big) \hat{q}+H.c.\Big) \Big\},
\label{eq:mc1rel}
\end{eqnarray}
where $\hat{\overline{G}}^{(k)}=\sum_{i=0,1,2}(G_{+,i}^{(k)}-G_{-,i}^{(k)})%
\hat{P}_{i}$ for $k=0,1$. The operator $\hat{W}_{0}(t)$ in Eq. (\ref
{eq:mc1rel}) corresponding to the zero order perturbation of noise spectrum
\cite{stace,li} is determined by the following master equation
\begin{eqnarray}
&&\frac{d}{dt}\hat{W}_{0}(t)=-i\hat{L}_{D}\hat{W}_{0}(t)-\lambda \Big\{%
\left[ \hat{q},\left[ \left[ \hat{G}^{(0)},\hat{W}_{0}(t)\right] \right]
\right]   \nonumber \\
&&~~~~~-\left( \left( \hat{G}^{(0)}\hat{\rho}_{0}(t)+2\hat{\overline{G}}%
^{(0)}\hat{N}_{0}(t)\right) \hat{q}+H.c.\right) \Big\}.  \label{eq:wl0}
\end{eqnarray}
In the derivation of Eqs. (\ref{eq:ml1rel},\ref{eq:mc1rel}), the ER decay
rates $\Gamma _{R,L}$ have been assumed to be independent of the tunneling
electron number $n$ for simplification. The detailed derivation of the above
master equations can be found in the Appendix.

As we can see, the EA effect is described by Eqs. (\ref{eq:ml1rel},\ref
{eq:mc1rel}). Eqs. (\ref{eq:ml0},\ref{eq:mc0}) are simply the rate equations
for the thermal equilibrium state without the EA effect.  If the bias
current flows from the source to the drain such that electrons accumulate in
the drain, the EA induces a negative effective bias voltage $-\delta V_{EA}$%
. Then, an additional current $I_{1}=\frac{d}{dt}$Tr$\left[ \xi \hat{N}%
_{1}\left( t\right) \right] $ is induced from the drain to the source. If
the bias current flows from the drain to the source such that electrons
accumulate in the source, a positive effective bias voltage $\delta V_{EA}$
is induced and an additional current flows from the source to the drain.
Eq.~(\ref{eq:mc1rel}) shows a modification of the first order perturbation
to the transport current in the QPC. Eqs. (\ref{eq:ml0}-\ref{eq:wl0}) are
the main result of the theory, which will be used to explore the quantum
decoherence of the measured qubit in the rest of the paper.

\section{DEPHASING AND RELAXATION IN THERMAL EQUILIBRIUM LIMIT}

As we have discussed in the previous section, the EA dynamics can be
considered as a perturbative modification of the thermal equilibrium
dynamics for the reduced system. Without considering the EA effect, the
electronic reservoir is treated as the thermal equilibrium state, in which
the ER process is assumed to respond fast and effectively. The measured
qubit in this case can be studied by the zero-order perturbation master
equation Eq. (\ref{eq:ml0}). When the EA effect also becomes significant,
the qubit undergoes a local non-equilibrium process initially, and then
approaches to a stable state asymptotically due to the bias effect. In this
section we will study the quantum decoherence of the qubit in the thermal
equilibrium limit. The EA effect on the qubit decoherence will be explored
in the next section.

In order to show the intrinsic measurement effect on the decoherence of the
qubit, we concentrate first on the qubit with symmetric CQD, namely, both
dots have equal energy levels $E_{R}=E_{L}$ and the qubit state is
characterized by $\theta =\pi /2$. The qubit with asymmetric CQD will be
studied later.

There have been several efforts contributed on this issues using the secular
approximation \cite{li} or discussed in the zero temperature limit \cite
{stace}. Here, an analytic discussion is presented. To discuss the dephasing
and the relaxation of the qubit in the measurement processes, a set of
matrix elements for the reduced density operator $\hat{\rho}_{0}$ are
introduced : $\rho _{0,r}=\rho _{0,ee}-\rho _{0,gg}$, $\rho _{0,d}=\rho
_{0,eg}+\rho _{0,ge}$ and $i\rho _{0,p}=\rho _{0,eg}-\rho _{0,ge}$, where $%
\rho _{0,ij}=\left\langle i\left| \hat{\rho}_{0}\right| j\right\rangle $,
and $\left| i,j=g\right\rangle $ ($\left| i,j=e\right\rangle $) are the
ground (excited) state of the qubit. Eq. (\ref{eq:ml0}) can then be
expressed as
\begin{eqnarray}
&&\dot{\rho}_{0,r}\left( t\right) =-\Gamma _{0,r}\left( \rho _{0,r}\left(
t\right) -\rho _{0,r}\left( \infty \right) \right) ,  \nonumber \\
&&\Gamma _{0,r}=\eta _{d}G_{+,a}^{\left( 0\right) },\;\rho _{0,r}\left(
\infty \right) =\frac{G_{+,b}^{\left( 0\right) }}{G_{+,a}^{\left( 0\right) }}%
,  \label{eq:l0relax}
\end{eqnarray}
and
\begin{equation}
\dot{\rho}_{0,d}\left( t\right) =\frac{\gamma }{\hbar }\rho _{0,p}\left(
t\right) ,\;\dot{\rho}_{0,p}\left( t\right) =-\frac{\gamma }{\hbar }\rho
_{0,d}\left( t\right) -\Gamma _{0,r}\rho _{0,p}\left( t\right) ,
\label{eq:l0deph}
\end{equation}
where $\eta _{d}=\pi g_{L}g_{R}\left( \delta \Omega \right) ^{2}/\hbar $,
and
\begin{eqnarray}
&&G_{+,a}^{(k)}=-\sum_{i=1,2}\left( \frac{G_{+,i}^{\left( k\right)
}+G_{-,i}^{\left( k\right) }}{2}\right) ,  \nonumber \\
&&G_{+,b}^{\left( k\right) }=\sum_{i=1,2}\left( -1\right) ^{i}\left( \frac{%
G_{+,i}^{\left( k\right) }+G_{-,i}^{\left( k\right) }}{2}\right) ,
\label{eq:locoeff}
\end{eqnarray}
with $k=0,1$.

\begin{figure}[tbph]
\includegraphics*[angle=90,scale=.35]{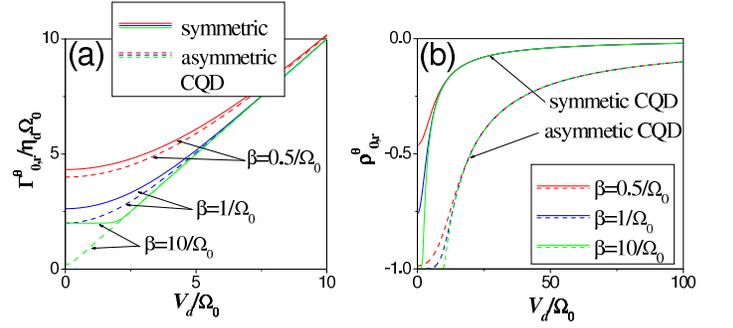}
\caption{(color online). (a) The decay rate $\Gamma _{0,r}^{\theta }$ of the
measured qubit in the equilibrium state. The solid curves are for symmetric
CQD $\left( \gamma =2\Omega _{0}\right) $ and the dash curves for asymmetric
CQD $\left( \gamma =10\Omega _{0}\right) $. (b) The asymptotic reduced
density matrix $\rho _{0,r}^{\theta }\left( \infty \right) $ of the measured
qubit. Two branches corresponds to the symmetric CQD (solid curves) and the
asymmetric CQD with $\gamma =10\Omega _{0}$(the dash curves), respectively.}
\label{fig2}
\end{figure}

According to Eq. (\ref{eq:l0relax}), the relaxation rate of the qubit is $%
\Gamma _{0,r}$. Applying the previous definitions in Eqs.
(\ref{eq:bg1}-\ref{eq:g01}), it leads to
\begin{equation}
\Gamma _{0,r}=\frac{\eta _{d}\left( V_{d}\sinh \beta V_{d}-\gamma \sinh
\beta \gamma \right) }{\cosh \beta V_{d}-\cosh \beta \gamma },\;\rho
_{0,r}\left( \infty \right) =\frac{-\gamma \eta _{d}}{\Gamma _{0,r}}.
\label{eq:monot}
\end{equation}
A plot of $\Gamma _{0,r}$ and $\rho _{0,r}\left( \infty \right) $ is shown
in Fig.~\ref{fig2}. The solid curves denote the symmetric CQD. The
relaxation rate $\Gamma _{0,r}$ is a positive-monotonic-increase function of
the temperature and bias voltage. Increasing temperature and bias voltage
will enhance qubit relaxation, as we expected. The asymptotic matrix element
$\rho _{0,r}\left( \infty \right) $ denotes an asymptotic stable state of
the qubit. It is a negative increase-monotonic function of temperature and
bias voltage with values in the range $-1\sim 0$. Because the external bias
can be regarded as a macroscopic reservoir (an effective heat bath) with
respect to the QPC, it is expected that there exists an analogy between
biased QPC and the heat bath \cite{mozyrsky}. A limited case is $\rho
_{0,r}\left( \infty \right) \rightarrow 0$ as $V_{d}>>\gamma $. The qubit
tends to stay in a completely random mixed state in the high bias limit.
This property is similar to the thermal randomness caused by the heat bath.
According to $\rho _{0,r}\left( \infty \right) $ in Eq. (\ref{eq:monot}),
the ground state occupation of the qubit shows a similar dependence on the
temperature effect and bias effect. Furthermore, it can be checked that $%
\Gamma _{0,r}\rightarrow \eta _{d}\gamma $ in the zero bias and zero
temperature limit. The relaxation due to QPC can not be stopped by turning
off bias. The QPC will exhaust energy and information of the qubit, as
pointed out first in \cite{mozyrsky}.

\begin{figure}[tbph]
\includegraphics*[angle=90,scale=.35]{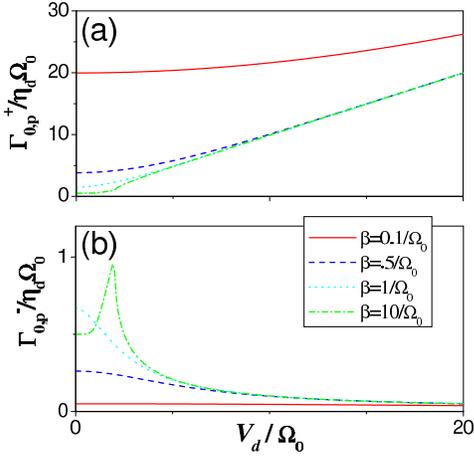}
\caption{(color online). (a) The dephasing rate $\Gamma _{0,p}^{+}$ of the
fast mode of the measured qubit in the equilibrium state. (b) The dephasing
rate $\Gamma _{0,p}^{-}$ of the slow mode for the measured qubit. The
figures are plotted with the fixed parameter $\gamma /\hbar \eta _{d}=\Omega
_{0}$.}
\label{fig3}
\end{figure}

The qubit dephasing is governed by the coupled differential equations (\ref
{eq:l0deph}), the solutions of which are a linear combination of two decay
modes
\begin{equation}
c_{1}e^{-\Gamma _{0,p}^{+}t}+c_{2}e^{-\Gamma _{0,p}^{-}t}
\label{eq:dephas_mode}
\end{equation}
with dephasing rates $\Gamma _{0,p}^{\pm }=\frac{1}{2}\Gamma _{0,r}\pm \frac{%
1}{2}\Gamma _{0,r}$ $\times \sqrt{1-\left( 2\gamma /\hbar \Gamma
_{0,r}\right) ^{2}}$, respectively. The mean dephasing rate is $\Gamma
_{0,r}/2$. The coefficients $c_{1,2}$ in Eq. (\ref{eq:dephas_mode}) depend
on the initial condition of the qubit. We then have the result
\begin{equation}
\rho _{0,p}\left( \infty \right) \rightarrow 0,\quad \rho _{0,d}\left(
\infty \right) \rightarrow 0.  \label{asy_pd}
\end{equation}
The qubit completely dephasing asymptotically. In addition, two decay modes
show different decay behavior in the regime $\hbar \Gamma _{0,r}/2>\gamma $.
A plot of $\Gamma _{0,p}^{\pm }$ versus $V_{d}$ is shown in Fig.~\ref{fig3}.
The curves with non-smooth turning point denote that $\Gamma _{0,p}^{\pm }$
has a transition from a degenerate regime $\hbar \Gamma _{0,r}/2<\gamma $,
in which both modes have the same dephasing rate $\Gamma _{0,r}/2$, to a
non-degenerate regime along $V_{d}$ axis. In the degenerate regime, only the
real part of $\Gamma _{0,p}^{\pm }$ is plotted. In Fig. \ref{fig3}a, the
fast decay mode, $\Gamma _{0,p}^{+}$, shows that the dephasing rate
increases as increasing temperature and bias voltage. In the high
temperature and high bias voltage limit, it approaches to the linear
dependence as the usual expectation. However, for the slow decay mode, $%
\Gamma _{0,p}^{-}$, Fig. \ref{fig3}b shows an upside-down dependence on the
temperature and bias voltage. At a given bias voltage, the dephasing rate of
the slow decay mode decreases with temperature increasing. It also shows
that the dephasing rate decreases with the bias voltage increasing at a
given temperature. The minimum dephasing rate appears in the large bias
voltage and large temperature limit. This phenomenon is very different from
the behavior of the fast decay mode that we have known. Even for the case of
high density of state (i.e. with small value of $\gamma /\hbar \eta _{d}$),
the phenomenon is still preserved. However, it should be noted that for a
quantum information process, quantum computations must deal with all
possible initial conditions. Therefore, the decoherence of the qubit is
characterized by the larger dephasing rate $\Gamma _{0,p}^{+}$ rather than
the slow one, $\Gamma _{0,p}^{-}$, unless an ancillary operation on the
qubit can be added to limit the initial conditions to a sub-domain belonging
to the slow decay mode.

Next we shall study the decoherence of the qubit with asymmetric CQD. In
this case, Eq. (\ref{eq:ml0}) can be expressed as
\begin{eqnarray}
&&\dot{\rho}_{0,d}(t)=\frac{\gamma }{\hbar }\rho _{0,p}(t)-\eta _{d}\cos
\theta \Big[\cos \theta G_{+}^{(0)}\rho _{0,d}(t)  \nonumber \\
&&~~~~~~~~~~~~~~~~~~+\sin \theta \Big(G_{+,b}^{(0)}-G_{+,a}^{(0)}\rho
_{0,r}(t)\Big) \Big],  \label{eq:asymast1} \\
&&\dot{\rho}_{0,r}(t)=\eta _{d}\sin \theta \Big[\cos \theta G_{+}^{(0)}\rho
_{0,d}(t)  \nonumber \\
&&~~~~~~~~~~~~~~~~~~+\sin \theta \Big(G_{+,b}^{(0)}-G_{+,a}^{(0)}\rho
_{0,r}(t)\Big) \Big],  \label{eq:asymast2} \\
&&\dot{\rho}_{0,p}(t)=\frac{-\gamma }{\hbar }\rho _{0,d}(t)-\eta _{d}\Big[%
G_{+}^{(0)}\cos \theta ^{2}  \nonumber \\
&&~~~~~~~~~~~~~~~~~~~~~~~~~~~~~~+G_{+,a}^{(0)}\sin \theta ^{2}\Big]\rho
_{0,p}(t),  \label{eq:asymast3}
\end{eqnarray}
where $G_{+}^{\left( 0\right) }=G_{+,0}^{\left( 0\right) }+G_{-,0}^{\left(
0\right) }=V_{d}\coth \left( \beta V_{d}/2\right) $. This is a set of
coupled differential equations. There is no analytic form to define the
decoherence rate like the case in the symmetric CQD. However, in the large
bias limit $\gamma \ll V_{d}$ we can find from the first two master
equations the relation $\dot{\rho}_{0,d}\left( t\right) /\dot{\rho}%
_{0,r}\left( t\right) \simeq -\cot \theta $. Then the relaxation rate can be
obtained
\begin{equation}
\Gamma _{0,r}^{\theta }=\eta _{d}\left( \cos \theta ^{2}G_{+}^{\left(
0\right) }+\sin \theta ^{2}G_{+,a}^{\left( 0\right) }\right) .
\label{eq:relaxasy}
\end{equation}
It can been shown that three decay rates from master equations (\ref
{eq:asymast1}-\ref{eq:asymast3}) correspond to the roots of the following
equation
\begin{equation}
0=y^{3}-2\Gamma _{0,r}^{\theta }y^{2}+\left( \frac{\gamma ^{2}+\left( \hbar
\Gamma _{0,r}^{\theta }\right) ^{2}}{\hbar ^{2}}\right) y-\eta
_{d}G_{+,a}^{\left( 0\right) }\left( \frac{\gamma \sin \theta }{\hbar }%
\right) ^{2}.  \label{eq:decayroot}
\end{equation}
The mean value of these roots is $2\Gamma _{0,r}^{\theta }/3$. $\Gamma
_{0,r}^{\theta }$ is also plotted in Fig.~\ref{fig2}a. All curves show a
positive-monotonic-increase with the temperature and bias voltage. This bias
and temperature dependence of the mean decay rate approaches to be linear in
the high bias regime. In the low temperature limit, it becomes linear for
almost the whole range of the bias voltage. In addition, it can be found in
Fig.~\ref{fig2}a that the asymmetric CQD has a longer decay time than the
symmetric CQD. It indicates that the influence of the device structure on
qubit decoherence can also become significant in the low temperature and low
bias regime.

The asymptotic behavior of Eqs. (\ref{eq:asymast1}-\ref{eq:asymast3}) can be
solved easily
\begin{eqnarray}
&& \rho _{0,d}^{\theta }\left( \infty \right) =\rho _{0,p}^{\theta }\left(
\infty \right) \rightarrow 0,  \label{eq:asymdp} \\
&& \rho _{0,r}^{\theta }\left( \infty \right) \rightarrow \frac{-\gamma
\left( \cosh \beta V_{d}-\cosh \beta \gamma \right) }{V_{d}\sinh \beta
V_{d}-\gamma \sinh \beta \gamma }.  \label{eq:asymr}
\end{eqnarray}
$\rho _{0,r}^{\theta }\left( \infty \right) $ implicitly depends on the
device structure through $\gamma $. The plot of $\rho _{0,r}^{\theta }\left(
\infty \right) $ in Fig.~\ref{fig2}b shows that there are two branches
corresponding to the symmetric (solid curves) and the asymmetric (dash
curves) CQD, respectively. At a given bias voltage, the symmetric CQD is
asymptotically forced into a more random mixed state in comparison with the
asymmetric CQD. This is because the device structure effectively modifies
the bias-dependence of the qubit coherence.

\section{EA EFFECT ON QUBIT}

Now, we turn to discuss the properties of the EA and the decoherence of the
charge qubit due to the EA effect.

\subsection{Properties of the EA}

\begin{figure}[tbph]
\includegraphics*[angle=90,scale=.6]{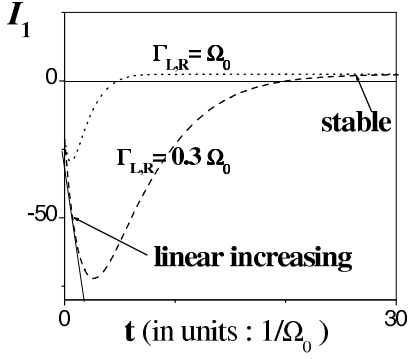}
\caption{The EA current $\emph{I}_{1}\left( t\right) $ in the QPC. The qubit
is initially in the $\left| L\right\rangle $ state without energy-level
offset. The following measurement parameters are used to plot figures : $%
\Omega =\Omega _{0}$, $\delta \Omega =0.1\Omega _{0}$, $\beta =1/\Omega _{0}$
, $\mu _{L}=100\Omega _{0}$, $(V_{d}=10\Omega _{0})$ and $g_{L,R}=2/\Omega
_{0}$. The initial input number of electrons in the source is set to be $100$%
. There is no electron accumulation at initial time. Two cases with
different ER decay rate are plotted.}
\label{fig4}
\end{figure}

The EA can be characterized by the EA induced current $I_{1}\left( t\right) =%
\frac{d}{dt}$Tr$\left[ \xi \hat{N}_{1}\left( t\right) \right] $. $%
I_{1}\left( t\right) $ is plotted in Fig.~\ref{fig4}, which is calculated
according to the resulted master equations. The result can be studied in
detail by the analysis of Eq. (\ref{eq:mc1rel}). In the right-hand side of
Eq. (\ref{eq:mc1rel}), because the traces of $-i\hat{L}_{D}\hat{N}_{1}\left(
t\right) $ and commutators vanish, the terms $\lambda \mathcal{F}\left(
t\right) \left( \hat{\overline{G}}^{(1)}\hat{N}_{0}\left( t\right) \hat{q}%
+H.c.\right) \equiv \hat{K}_{1}$ and $\lambda \left( \hat{\overline{G}}^{(0)}%
\hat{\rho}_{1}(t)\hat{q}+H.c.\right) \equiv \hat{K}_{0}$ completely
determine the EA current. Also, the factor $\mathcal{F}\left( t\right) $ of $%
\hat{K}_{1}$ is exponential decay in time, and $\hat{\rho}_{1}\left(
t_{0}\right) \ll \hat{N}_{0}\left( t_{0}\right) $. Thus, $\hat{K}_{1}$ is
initially dominant, and then $\hat{K}_{0}$ becomes significant as $\hat{K}%
_{1}$ exponentially decays to zero. The linear increase of the EA current in
the beginning is governed mainly by $\hat{K}_{1}$, see Fig.~{\ref{fig4}. }

Analytically, $\hat{N}_{0}\left( t\right) $ can be solved in the asymptotic
limit
\begin{eqnarray}
&&n_{0,gg}\left( t\right) =c_{g0}+c_{g1}t,\;n_{0,ee}\left( t\right)
=c_{e0}+c_{e1}t,  \nonumber \\
&&c_{e1}=c_{0}\left( G_{+,1}^{\left( 0\right) }+G_{-,1}^{\left( 0\right)
}\right) ,\;c_{g1}=c_{0}\left( G_{+,2}^{\left( 0\right) }+G_{-,2}^{\left(
0\right) }\right) ,  \nonumber \\
&&  \label{eq:n1slope}
\end{eqnarray}
where $c_{g0}$ and $c_{e0}$ are simply constants, $c_{0}=-\lambda \Omega
^{2}V_{d}/G_{+,a}^{\left( 0\right) }$, the higher order contributions of $%
\delta \Omega $ to the coefficient $c_{0}$ have also been ignored. For the
weak interaction coupling, it can be numerically checked that $%
n_{0,ge}\left( t\right) \ll n_{0,gg(ee)}\left( t\right) $ in the asymptotic
limit. We then obtain
\begin{equation}
\text{Tr}\left[ \hat{K}_{1}\right] =-\mathcal{F}\left( t\right) \left(
\left( 2\lambda \Omega ^{2}\right) ^{2}V_{d}t+\text{constant}\right) .
\label{eq:k1asympt}
\end{equation}
In the beginning (the $\hat{K}_{1}$ dominant regime), the ER decay is
adiabatic, namely, $\mathcal{F}\left( t\right) \approx $ constant ($\equiv
\mathcal{F}^{\prime }$). It leads to
\begin{equation}
I_{1}\left( t\right) \approx -\mathcal{F}^{\prime }\xi \left( 2\lambda
\Omega ^{2}\right) ^{2}V_{d}t+\text{constant}.  \label{eq:ea1vnon}
\end{equation}
This shows why the EA current is linear increasing in time, as one can see
in Fig.~\ref{fig4}.

In the crossover regime in Fig.~\ref{fig4}, the EA is slowed down by the
bias. The maximum amount of accumulated electrons occurs at the valley of
the $I_{1}$-curves, in which the maximum reverse EA current is induced.
Obviously, the smaller the ER decay rate $\Gamma _{R,L}$, the stronger the
EA as shown in Fig.~\ref{fig4}.

Due to the bias, electrons accumulated in the QPC are completely exhausted
through the ER process. Tr$\left[ \hat{K}_{1}\right] $ of Eq. (\ref
{eq:mc1rel}) vanishes in this stage, and Tr$\left[ \hat{K}_{0}\right] $
becomes dominant. The asymptotic current can then be solved analytically
\begin{eqnarray}
I_{1}(\infty ) &=&\frac{d}{dt}\text{Tr}\left[ \xi \hat{N}_{1}(t)\right] =\xi
\text{Tr}[\hat{K}_{0}(\infty )]  \nonumber \\
&=&2\lambda \xi V_{d}\left( \Omega ^{2}-\Omega \delta \Omega \left( 1+\rho
_{1,r}\left( \infty \right) \cos \theta \right) \right) ,  \nonumber \\
&&  \label{eq:cur1asy}
\end{eqnarray}
where the higher order contribution from $\delta \Omega ^{2}$ has been
ignored, and the first-order relaxation shift $\rho _{1,r}\left( \infty
\right) $ is given in Eq. (\ref{eq:l1rasympt}). It is worth noting that the
qubit density $\rho _{1,r}\left( t\right) $ records the history of the EA
effect. Even the accumulated electrons has been exhausted in this stage, $%
\rho _{1,r}\left( t\right) $ in $\hat{K}_{0}(t)$ performs as an EA
background field and affects the electron tunneling in the QPC. Also, the
time independent relaxation shift $\rho _{1,r}\left( \infty \right) $ leads
to a constant transmission probability for the first-order electron
tunneling. Finally, the stable behavior of the EA current is reached, see
Fig.~\ref{fig4}. As a result, with the EA effect, the asymptotic current up
to the first order perturbation is
\begin{eqnarray}
&&I_{0}(\infty )+I_{1}(\infty )=2\lambda \left( \Omega -\delta \Omega
\right) ^{2}V_{d}  \nonumber \\
&&~~~~~~~~~~~~~~~~~~~~+\frac{\lambda \delta \Omega ^{2}}{2}\frac{\left(
V_{d}^{2}-\gamma ^{2}\right) \sinh \beta V_{d}}{V_{d}\sinh \beta
V_{d}-\gamma \sinh \beta \gamma }  \nonumber \\
&&~~~~~~~~~ +2\lambda \xi V_{d}\left( \Omega ^{2}-\Omega \delta \Omega
\left( 1+\rho _{0,r}^{\theta }\left( \infty \right) \cos \theta \right)
\right).  \nonumber \\
&&
\end{eqnarray}

\subsection{Decoherence induced by EA effect}

\begin{figure}[tbph]
\includegraphics*[angle=90,scale=.55]{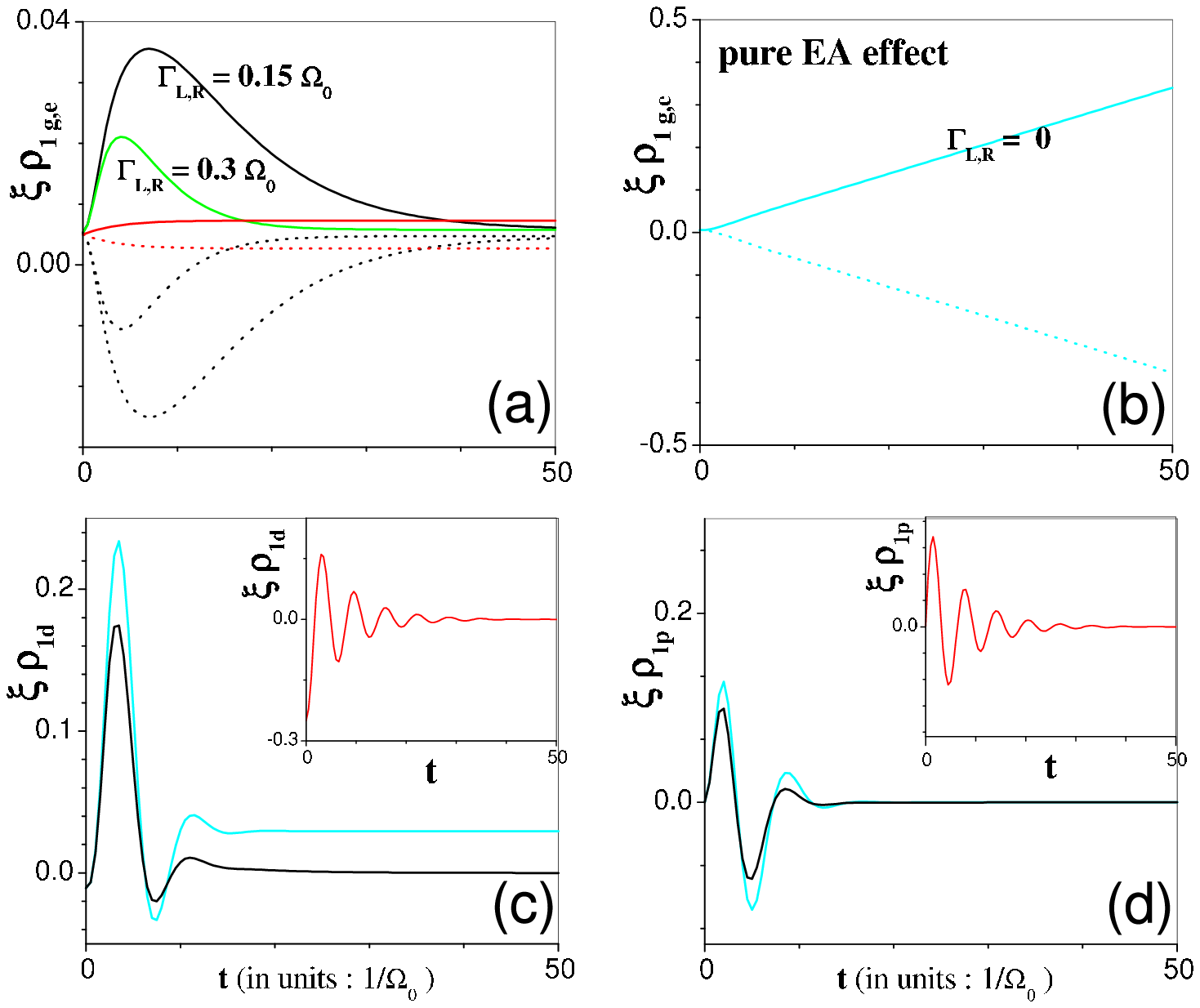}
\caption{(color online). The time evolution of the reduced density matrix of
the first order perturbation. The symmetric CQD is simulated. The qubit is
initially in $\left| L\right\rangle $ state with $\xi \mathtt{Tr}[\hat{\rho}%
_{1}\left( t=0\right) ]=0.01 $. The input initial electron number is $100$.
The following measurement parameters are used : $\Omega =\Omega _{0}$, $%
\delta \Omega =0.1\Omega _{0}$, $\beta =1/\Omega _{0}$ , $\mu _{L}=100\Omega
_{0}$, $g_{L,R}=2/\Omega _{0}$, $\omega _{d}=0$ and $\mu _{R}=\mu _{L}-V_{d}$
with $V_{d}=0$ and $\Gamma _{L,R}=0.15\Omega _{0}$ for red curve, $%
V_{d}=10\Omega _{0}$ and $\Gamma _{L,R}=0.15\Omega _{0}$ black curve, $%
V_{d}=10\Omega _{0}$ and $\Gamma _{L,R}=0.3\Omega _{0} $ green curve, and $%
V_{d}=10\Omega _{0}$ and $\Gamma _{L,R}=0$ cyan curve, respectively. (a) The
qubit relaxation. The time evolution of the matrix elements $\rho _{1g}$ $%
\left( \rho _{1e}\right) $ for the qubit in the ground (excited) state are
plotted by solid (dash) curves, respectively. (b) The qubit relaxation under
the pure EA effect ($\Gamma _{L,R}=0$). (c) and (d) The qubit dephasing. The
time evolution of the matrix element $\rho _{1d}$ and $\rho _{1p}$ are
plotted, respectively.}
\label{fig5}
\end{figure}

The qubit relaxation and dephasing due to the EA effect are studied in this
subsection. Eq.~(\ref{eq:ml1rel}) describes this decoherence process. The
terms $\lambda \mathcal{F}\left( t\right) \left[ \left[ \left[ \hat{G}%
^{\left( 1\right) },\hat{N}_{0}\left( t\right) \right] \right] ,\hat{q}%
\right] \equiv \hat{K}_{1}^{\prime }$ in Eq. (\ref{eq:ml1rel}) is related to
$\hat{K}_{1}$ of Eq. (\ref{eq:mc1rel}) through the relation $\hat{N}_{k}={%
\sum_{n=0}^{\infty }}n\hat{\rho}_{k,f}^{(n)}(t)-{\sum_{m=0}^{\infty }}m\hat{%
\rho}_{k,b}^{(m)}(t)$. This term arises from the EA, and the induced
relaxation is suppressed by the bias. To understand this property, let us
look at the asymptotic matrix elements $\left\langle i\left| \hat{K}%
_{1}^{\prime }(\infty )\right| i\right\rangle $ first. Applying the previous
result Eq. (\ref{eq:n1slope}) and the asymptotic analysis, we obtain, up to
the order $\delta \Omega $,
\begin{eqnarray}
&&\left\langle g\left| \hat{K}_{1}^{\prime }(\infty )\right| g\right\rangle
=K_{g0},  \nonumber \\
&&\left\langle e\left| \hat{K}_{1}^{\prime }(\infty )\right| e\right\rangle
=-\left\langle g\left| \hat{K}_{1}^{\prime }(\infty )\right| g\right\rangle ,
\label{eq:kg0}
\end{eqnarray}
where
\[
K_{g0}\approx \lambda \Omega \delta \Omega \mathcal{F}\left( t\right) V_{d}%
\frac{(G_{+,b}^{\left( 0\right) }-G_{+,a}^{\left( 0\right)
})(G_{+,a}^{\left( 1\right) }+G_{+,b}^{\left( 1\right) })}{(G_{+,a}^{\left(
0\right) })^{2}},
\]
and $G_{+,(a,b)}^{\left( 0,1\right) }$ is defined in Eq.~(\ref{eq:locoeff}).
It can be checked that $K_{g0}$ is positive because of $(G_{+,b}^{\left(
0\right) }-G_{+,a}^{\left( 0\right) })\leq 0$ and $-1<(G_{+,a}^{\left(
1\right) }+G_{+,b}^{\left( 1\right) })\leq 0$. Since $\hat{\rho}_{1}\left(
t_{0}\right) \ll \hat{N}_{0}\left( t_{0}\right) $, $\frac{d}{dt}\left\langle
i\left| \hat{\rho}_{1}(t)\right| i\right\rangle \approx \left\langle i\left|
\hat{K}_{1}^{\prime }\right| i\right\rangle $ before the bias effect becomes
active (i.e. in the EA effect dominant regime). We then have
\begin{eqnarray}
\left\langle g\left| \hat{\rho}_{1}\left( t\right) \right| g\right\rangle
&\approx &\left. K_{g0}\right| _{\mathcal{F}=1}t\text{ + constant},
\nonumber \\
\left\langle e\left| \hat{\rho}_{1}\left( t\right) \right| e\right\rangle
&\approx &-\left. K_{g0}\right| _{\mathcal{F}=1}t\text{ + constant}.
\label{eq:relax}
\end{eqnarray}
This linear time-dependence behavior indeed indicates an extra \textit{qubit
relaxation} due to the EA effect. The numerical results in Fig. ~\ref{fig5}a
(plotted by green and black curves) roughly show this linear time-dependence
of the relaxation. In Fig. \ref{fig5}a, the solid (dash) lines correspond to
the ground (excited) state, respectively. Instructively, a numerical result
for $\Gamma _{L,R}=0$ is plotted in Fig.~\ref{fig5}b. A perfect linear
character is obtained. The induced relaxation from the \textit{pure} EA
effect obeys the linear time dependence, as shown by Eq. (\ref{eq:relax}).

According to Eq.~(\ref{eq:kg0}), it is obvious that this relaxation process
is slowed down by the bias because $K_{g0}$ reduces to zero by the decay
factor $\mathcal{F}(t)$. This slow-down effect causes the qubit decay to the
lowest energy state (in the first order perturbation) in the relaxation
process, see Fig.~\ref{fig5}a. Then, the term $\lambda \left[ \left[ \left[
\hat{G}^{(0)},\hat{\rho}_{1}(t)\right] \right] ,\hat{q}\right] \equiv \hat{K}%
_{0}^{\prime }$ of Eq. (\ref{eq:ml1rel}) becomes dominant. The qubit state
undergoes a transition from the relaxation to excitation, see Fig.~\ref{fig5}%
a. After the accumulated electrons exhausted, this excitation is completely
governed by $\hat{K}_{0}^{\prime }$. In addition, the black and green curves
in Fig.~\ref{fig5}a show that a stronger EA effect (i.e. with smaller ER
decay rate) forces the qubit to experience a larger relaxation shift (i.e. a
lower energy state). The asymptotic relaxation shift $\rho _{1,r}(\infty) $
can be easily solved
\begin{eqnarray}
\rho _{1,r}\left( \infty \right) &=&\left\langle e\left| \hat{\rho}%
_{1}\left( \infty \right) \right| e\right\rangle -\left\langle g\left| \hat{%
\rho}_{1}\left( \infty \right) \right| g\right\rangle  \nonumber \\
&\rightarrow &\frac{-\gamma \left( \cosh \beta V_{d}-\cosh \beta \gamma
\right) }{V_{d}\sinh \beta V_{d}-\gamma \sinh \beta \gamma },
\label{eq:l1rasympt}
\end{eqnarray}
which is independent of the ER decay rate. As a result, the total qubit
relaxation up to the first order perturbation is
\begin{eqnarray}
&&\rho _{tot,r}\left( \infty \right) =\frac{-\gamma \left( \cosh \beta
V_{d}-\cosh \beta \gamma \right) }{V_{d}\sinh \beta V_{d}-\gamma \sinh \beta
\gamma }  \nonumber \\
&&\times \Big[1+\frac{2}{\left( \mu _{L}+\mu _{R}\right) }\left( \frac{%
1+e^{-\beta \mu _{L}}}{Ag_{L}}+\frac{1+e^{-\beta \mu _{R}}}{Ag_{R}}\right) %
\Big].  \nonumber \\
&&
\end{eqnarray}
The perturbed term comes from the EA effect.

We may also discuss the low bias voltage limit, $V_{d}\ll \gamma $. It can
be checked that $\hat{G}^{\left( 1\right) }\rightarrow 0$ as $%
V_{d}\rightarrow 0$. Thus, $\hat{K}_{1}^{\prime }$ in Eq. (\ref{eq:ml1rel})
vanishes, and the EA makes almost no effect on the qubit dynamics. The
result is shown by the red curve in Fig.~\ref{fig5}a. The qubit only
experiences a relaxation process without the excitation one.

Next, we shall calculate the decoherence rate of the qubit. For a low bias
voltage, because of $\hat{G}^{\left( 1\right) }\rightarrow 0$ as $%
V_{d}\rightarrow 0$, the dynamic structure of Eq. (\ref{eq:ml0}) and (\ref
{eq:ml1rel}) are almost the same. The decay modes of the qubit contributed
by the zero order and the first order perturbation are closer. The EA effect
can only enhance the qubit relaxation. Dynamically, the decoherence rate of
the qubit does not be speeded up by the EA effect. According to Eq. (\ref
{eq:monot}), we obtain the total relaxation time $T_{tot,r}$ and the
dephasing time $T_{tot,p}$ of the qubit with the EA effect
\begin{equation}
T_{tot,r}=\frac{1}{\left. \Gamma _{0,r}\right| _{V_{d}\rightarrow 0}}=\frac{%
\cosh \beta \gamma -1}{\eta _{d}\gamma \sinh \beta \gamma }%
,\;T_{tot,p}=2T_{tot,r}  \label{eq:dtsym}
\end{equation}
for the symmetric CQD in the low bias limit.

With bias voltage increasing, the first-order qubit dephasing rate can not
be solved exactly. We study the qubit dephasing according to Eq. (\ref
{eq:ml1rel}), or in terms of the set of coupled differential equations for
the symmetric CQD
\begin{eqnarray}
\dot{\rho}_{1,d}\left( t\right) &=&\frac{\gamma }{\hbar }\rho _{1,p}\left(
t\right) ,  \nonumber \\
\dot{\rho}_{1,p}\left( t\right) &=&-\frac{\gamma }{\hbar }\rho _{1,d}\left(
t\right) -\Gamma _{0,r}\rho _{1,p}\left( t\right)  \nonumber \\
&&~~~~~~~-\Gamma _{1,r}\mathcal{F}\left( t\right) n_{0,p}\left( t\right) ,
\label{eq:1dephas}
\end{eqnarray}
where
\begin{equation}
\Gamma _{1,r}=-\eta _{d}G_{+,a}^{\left( 1\right) },  \label{eq:iearr}
\end{equation}
$\rho _{1,d}=\rho _{1,eg}+\rho _{1,ge}$, $i\rho _{1,p}=\rho _{1,eg}-\rho
_{1,ge}$ and $in_{0,p}=n_{0,ge}-n_{0,eg}$. The term $\Gamma _{0,r}\rho
_{1,p}\left( t\right) $ in Eq. (\ref{eq:1dephas}) denotes the qubit
dephasing arisen from the QPC-dots interaction associated with the bias,
while the term $\Gamma _{1,r}\mathcal{F}\left( t\right) n_{0,p}\left(
t\right) $ is resulted from the EA. Because $\Gamma _{1,r}$ is proportional
to the mean chemical potential $\bar{\mu}$ $(=\frac{\mu _{L}+\mu _{R}}{2})$%
,\ $\Gamma _{1,r}$ can be larger than $\Gamma _{0,r}$ for $\bar{\mu}%
>V_{d}\gg \gamma $. The qubit dephasing is mainly determined by the ER decay
rate and the mean chemical potential, it does not so sensitively
depend on the bias voltage. However, in the low bias limit,
especially for $V_{d}\ll \gamma $, $G_{+,a}^{\left( 1\right)
}\rightarrow 0$ (i.e. $\Gamma _{1,r}\rightarrow 0$). The qubit
dephasing rate becomes much smaller. The numerical plot of the
time evolution of the qubit off-diagonal matrix element shows the
coincident result, as one can see in Fig.~\ref{fig5}c and d.

\subsection{Decoherenceless EA effect in the low temperature regime}

\begin{figure}[tbph]
\includegraphics*[angle=90,scale=.6]{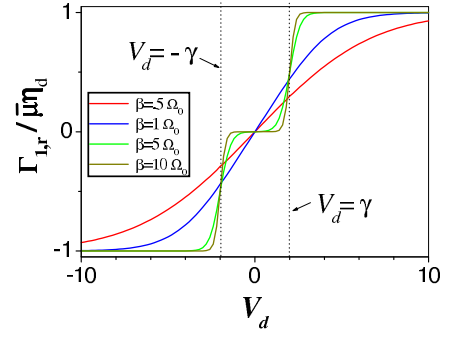}
\caption{(color online). The EA induced relaxation rate. The curves are
plotted with different temperatures.}
\label{fig6}
\end{figure}

We have pointed out that the qubit undergoes an extra relaxation under the
EA effect. However, this EA induced relaxation can be suppressed in the low
temperature regime. The symmetric CQD is used to check this result. The
qubit relaxation rate in the first order perturbation can be deduced from
Eq. (\ref{eq:ml1rel})
\begin{equation}
\Gamma _{1,r}(t)\equiv \frac{-\dot{\rho}_{1,r}\left( t\right) }{(\rho
_{1,r}\left( t\right) -\rho _{1,r}\left( \infty \right) )}=\Gamma _{0,r}+%
\mathcal{K}\left( t\right) \Gamma _{1,r}
\end{equation}
where $\mathcal{K}\left( t\right) =\mathcal{F}\left( t\right) n_{0,r}\left(
t\right) /(\rho _{1,r}\left( \infty \right) -\rho _{1,r}\left( t\right) )$,
and $n_{0,r}\left( \infty \right) =n_{0}G_{+,b}^{\left( 1\right)
}/G_{+,a}^{\left( 1\right) }=0$ has been used ($n_{0}\equiv $Tr$\left[ \hat{N%
}_{0}(t_{0})\right] $), and $\Gamma _{1,r}$ is defined in Eq. (\ref{eq:iearr}%
). The relaxation rate $\Gamma _{1,r}(t)$ is time dependent. Due to the EA
effect, initially, the qubit severely relaxes with the relaxation rate $%
\Gamma _{1,r}(t)\approx \Gamma _{1,r}\mathcal{K}\left( t\right) $ for $\bar{%
\mu}\gg V_{d}$. After the EA effect suppressed by the bias, the relaxation
rate asymptotically reduces to the zero-order $\Gamma _{0,r}$. Here, $%
\mathcal{K}\left( t\right) $ is a fluctuant factor. The EA induced
relaxation rate $\Gamma _{1,r}$ is plotted in Fig.~\ref{fig6}. It shows that
$\Gamma _{1,r}$ has a very different behavior from $\Gamma _{0,r}$. For a
large mean chemical potential ($\bar{\mu}\gg V_{d}$) in which $\bar{\mu}$ is
almost bias voltage independent, the electron source of the EA is excited
from the energy levels below the Fermi surface. The variation of the
chemical potential $\delta V_{EA}$ induced by the EA is limited by the
chemical potential. The $\delta V_{EA}$-dependence of the qubit relaxation
rate leads to a bounded phenomenon, as shown in Fig.~\ref{fig6}.

The relaxation rate $\Gamma _{1,r}$ also shows an anti-symmetric bias
dependence (see Fig.~\ref{fig6}). This feature can be understood as follows.
There are two kind of electron-tunneling correlations, $G_{+,i}^{(k)}$ and $%
G_{-,i}^{(k)}$, are involved. $G_{+,i}^{(k)}$ ($G_{-,i}^{(k)}$) corresponds
to the electron tunneling that one electron is from the source to the drain
at $\tau $ $(t)$ and the other is from the drain to the source at $t$ $(\tau
)$. Explicitly, $G_{+(-),i}^{(k)}$ is associated with $\mathrm{Tr}_{D^{n}}[%
\hat{f}_{t}^{+}\hat{f}_{\tau }\hat{\rho}_{I}(t)]$ $(\mathrm{Tr}_{D^{n}}[\hat{%
f_{t}}\hat{f}_{\tau }^{+}\hat{\rho}_{I}(t)])$ for forward tunneling
processes and $\mathrm{Tr}_{\bar{D}^{m}}[\hat{f_{t}}^{+}\hat{f}_{\tau }\hat{%
\rho}_{I}(t)]$ $(\mathrm{Tr}_{\bar{D}^{m}}[\hat{f_{t}}\hat{f}_{\tau }^{+}%
\hat{\rho}_{I}(t)])$ for backward tunneling processes. The relaxation rate
is related to $G_{+,i}^{(k)}$ and $G_{-,i}^{(k)}$ through the relation $%
\Gamma _{k,r}=\eta _{d}\sum_{i=1,2}\left( G_{+,i}^{\left( k\right)
}+G_{-,i}^{\left( k\right) }\right) /2$ with $k=0,1$. For $k=0$, the
electron-tunneling correlation is simply governed by the QPC-dots
interaction and the bias, no EA involved. It can be checked that a relation
of the bias symmetry holds
\begin{equation}
\frac{G_{+,i}^{\left( 0\right) }(V_{d})+G_{-,i}^{\left( 0\right) }(V_{d})}{2}%
=\frac{G_{+,i}^{\left( 0\right) }(-V_{d})+G_{-,i}^{\left( 0\right) }(-V_{d})%
}{2}.
\end{equation}
The $\Gamma _{0,r}$ does not be changed by reversing the bias ($%
V_{d}\rightarrow -V_{d}$), due to the fact that the qubit relaxation caused
by the electron tunneling in the QPC only depends on the \textit{amplitude}
of the external bias voltage. However, the electron-tunneling correlation
under the EA effect $(k=1)$ obeys the relation of the bias anti-symmetry
\begin{equation}
\frac{G_{+,i}^{\left( 1\right) }(V_{d})+G_{-,i}^{\left( 1\right) }(V_{d})}{2}%
=-\frac{G_{+,i}^{\left( 1\right) }(-V_{d})+G_{-,i}^{\left( 1\right) }(-V_{d})%
}{2}.  \label{antisym}
\end{equation}
To understand this relation, let us look at the forward tunneling processes,
in which an extra negative bias voltage is induced by the EA, $%
V_{d}\rightarrow V_{d}-\delta V_{EA}$. According to the result in Sec. IV B,
an effective \textit{relaxation} of the qubit is induced by the EA effect.
On the other hand, for a negative bias voltage, an extra negative bias
voltage induced by the EA leads to $-V_{d}\rightarrow -V_{d}-\delta V_{EA}$.
The amplitude of bias voltage is increased. An effective \textit{excitation}
of the qubit is induced by the EA effect. Accordingly, Eq. (\ref{antisym})
indicates that the EA induced relaxation rate of the qubit under the
positive bias voltage ($V_{d}$) is equal to the EA induced excitation rate
of the qubit under the negative bias voltage ($-V_{d}$).

In addition, the relaxation process can be separated into two effective
modes
\begin{equation}
\Gamma _{1,r}=\eta _{d}\bar{\mu}(\frac{R(z_{+})+R(z_{-})}{2}),\quad z_{\pm }=%
\frac{\beta (V_{d}\pm \gamma )}{2},
\end{equation}
where the function $R(z)=\coth (z)-z\csc $h$(z)^{2}$ is a step-like function
with bounded values $\pm \frac{1}{2}$. The function preserves the bias
anti-symmetry. As discussed in Sec. III, the relaxation depends on the
structure of the CQD. It leads to two effective modes corresponding to $%
R(z_{+})$ and $R(z_{-})$. In the low temperature regime, these two effective
modes separate. Eq. (\ref{antisym}) shows that the qubit excitation rate in
the $R(z_{-})$ mode cancels the qubit relaxation rate in the $R(z_{+})$ mode
in the regime of $V_{d}$ limited by $\pm \gamma $ . Then a relaxationless EA
effect occurs, as shown in Fig.~\ref{fig6}. However, in the high temperature
regime, the CQD-structure dependence of the relaxation is suppressed, and
two modes $R(z_{\pm })$ become degenerate. No relaxationless EA effect
occurs. As a result, $\Gamma _{1,r}(t)$ reduces to $\Gamma _{0,r}$ (i.e. $%
\Gamma _{1,r}\rightarrow 0$) in low temperature regime. For the qubit
dephasing, it can be checked that the dephasing term $\Gamma _{1,r}\mathcal{F%
}\left( t\right) n_{0,p}\left( t\right) \rightarrow 0$ as $T\rightarrow 0$.
The qubit is also EA-dephasingless in the low temperature regime.

\section{SUMMARY}

We have developed a perturbative theory to study the EA effect on the
decoherence of the charge qubit, a single electron confined in CQD, measured
by the QPC. The contribution due to the EA is treated perturbatively. A set
of master equations for the reduced density matrix of the qubit has been
obtained in this perturbation scheme. By solving the resulted master
equations, we obtain the decay modes of the dephasing and the relaxation of
the qubit in the thermal equilibrium limit, and study the temperature- and
bias-dependence of the dephasing and the relaxation rate. We find two kinds
of decay modes for the dephasing in the electrical measurement processes:
one decay rate increases as the increase of temperature and bias voltage, as
expected; the other shows an abnormal dependence. The qubit in the later
decay mode preserves the longest decoherence time in the high temperature
and high bias limit, its time scale is much longer than that of the usual
mode. In addition, the EA properties are studied extensively. The EA current
is obtained according to the master equations. We find that the EA current
varies linearly with time under the pure EA effect, and is then slowed down
into an asymptotic stable state by the bias. The qubit decoherence due to
the EA effect is studied based on this analysis. We find that under the EA
effect the qubit relaxation and dephasing rate are much larger than the ones
in the thermal equilibrium limit, unless the bias voltage is turned off.
Also, the master equations show that an extra qubit relaxation is induced by
the EA effect initially, and then be suppressed by the bias. Asymptotically,
the qubit will be forced into the state with a small relaxation shift which
is independent of the ER decay rate. However, in the low bias limit, the
qubit will not be affected by the EA. The qubit decoherence rate will not be
speeded up by the EA effect. Finally, we find a decoherenceless EA effect in
the low temperature regime. A bias anti-symmetry in EA processes suppresses
the qubit decoherence. These decoherent behaviors worth being explored in
experiments.

\appendix*

\section{DERIVATION OF MASTER EQUATIONS}

The derivation of the master equations (\ref{eq:ml0},\ref{eq:ml1rel},\ref
{eq:mc0},\ref{eq:mc1rel}) is presented in this appendix. The total density
operator in the interaction picture of the reservoir satisfies
\begin{equation}
\frac{d}{dt}\hat{\rho}_{I}\left( t\right) =\frac{1}{i\hbar }\left[ \hat{H}%
_{S}+\hat{H}_{I}^{\prime },\hat{\rho}_{I}\left( t\right) \right] ,\;\hat{H}%
_{I}^{\prime }=e^{\frac{i\hat{H}_{B}t}{\hbar }}\hat{H}^{\prime }e^{\frac{%
\hat{H}_{B}t}{i\hbar }}.  \label{eq:A1}
\end{equation}
In the weak coupling regime, the interaction Hamiltonian $\hat{H}^{\prime }$
can be treated by using the second order cummulant expansion technique \cite
{mori,li,goan,stace}. By taking the conditional trace Tr$_{D^{\left(
n\right) }\left( \bar{D}^{\left( m\right) }\right) }$ shown in Sec. II and
the second order cummulant expansion, the master equation becomes
\begin{eqnarray}
&&\frac{d}{dt}\hat{\rho}_{f\left( b\right) }^{\left( n\right) }\left(
t\right) =-i\hat{L}_{D}\hat{\rho}_{f\left( b\right) }^{\left( n\right)
}\left( t\right) -\hat{R}_{f\left( b\right) }\hat{\rho}_{f\left( b\right)
}^{\left( n\right) }\left( t\right) ,  \label{eq:A2} \\
&&\hat{R}_{f\left( b\right) }\hat{\rho}_{f\left( b\right) }^{\left( n\right)
}\left( t\right) =\frac{1}{\hbar ^{2}}\int_{t_{0}}^{t}d\tau \text{Tr}%
_{D^{(n)}(\bar{D}^{(n)})}  \nonumber \\
&&~~~~~~\times \left\{ \Big[\hat{H}_{I}^{\prime }(t),\left[ \hat{G}\left(
t,\tau \right) \hat{H}_{I}^{\prime }(\tau )\hat{G}\left( t,\tau \right) ^{+},%
\hat{\rho}_{I}\left( t\right) \right] \Big]\right\} ,  \nonumber \\
&&  \label{eq:A3}
\end{eqnarray}
where $\hat{\rho}_{f\left( b\right) }^{\left( n\right) }\left( t\right) =%
\mathrm{Tr}_{D^{(n)}(\bar{D}^{(n)})}[\hat{\rho}_{tot}(t)]$ and $\hat{G}%
\left( t,\tau \right) $ the propagator of the CQD. The decoherence
of the measured qubit is governed by the dissipation term
$-\hat{R}_{f\left( b\right) }\hat{\rho}_{f\left( b\right)
}^{\left( n\right) }\left( t\right) $ in Eq. (\ref{eq:A2}).

Note that the master equations (\ref{eq:A2}, \ref{eq:A3})
essentially describe a Markovian process. The Markovian
approximation has been introduced in this stage. Therefore, the
time integration in Eq. (\ref{eq:A3}) is replaced by the one be
integrated out along the whole time domain. Carrying out the commuter in Eq. (\ref{eq:A3}), $\hat{R}_{f\left( b\right) }%
\hat{\rho}_{f\left( b\right) }^{\left( n\right) }\left( t\right) $ can be
expressed as
\begin{eqnarray}
&&\frac{1}{2 \hbar ^{2}}\int_{-\infty}^{\infty}d\tau \Big\{\hat{q}(\hat{P}_{0}-e^{%
\frac{i\gamma \left( \tau -t\right) }{\hbar
}}\hat{P}_{1}-e^{\frac{-i\gamma
\left( \tau -t\right) }{\hbar }}\hat{P}_{2})  \nonumber \\
&&~~~~~~~\times (C_{f(b),-}^{n}(t-\tau )+C_{f(b),+}^{n}(t-\tau ))\Big\}+H.c.
\nonumber \\
&&-\frac{1}{2 \hbar ^{2}}\int_{-\infty}^{\infty}d\tau \Big\{(\hat{P}_{0}-e^{\frac{%
i\gamma \left( \tau -t\right) }{\hbar
}}\hat{P}_{1}-e^{\frac{-i\gamma \left(
\tau -t\right) }{\hbar }}\hat{P}_{2})  \nonumber \\
&&~~~~~~~\times (C_{f,+(b,-)}^{n-1}(t-\tau )+C_{f,-(b,+)}^{n+1}(t-\tau ))%
\hat{q}\Big\}+H.c.,  \nonumber \\
&&  \label{eq:A4}
\end{eqnarray}
where the reservoir time correlation functions are defined by
\begin{eqnarray}
C_{f,+}^{n}(t-\tau ) &=&\mathrm{Tr}_{D^{(n)}}[\hat{f}_{t}^{+}\hat{f}_{\tau }%
\hat{\rho}_{I}(t)],  \nonumber \\
C_{f,-}^{n}(t-\tau ) &=&\mathrm{Tr}_{D^{(n)}}[\hat{f_{t}}\hat{f}_{\tau }^{+}%
\hat{\rho}_{I}(t)],  \label{eq:c1}
\end{eqnarray}
for the forward tunneling processes, and
\begin{eqnarray}
C_{b,+}^{m}(t-\tau ) &=&\mathrm{Tr}_{\bar{D}^{(m)}}[\hat{f}_{t}^{+}\hat{f}%
_{\tau }\hat{\rho}_{I}(t)],  \nonumber \\
C_{b,-}^{m}(t-\tau ) &=&\mathrm{Tr}_{\bar{D}^{(m)}}[\hat{f_{t}}\hat{f}_{\tau
}^{+}\hat{\rho}_{I}(t)],  \label{eq:c2}
\end{eqnarray}
for the backward tunneling processes. It can be easily checked from Eqs.~(%
\ref{eq:c1},\ref{eq:c2}) that without the EA effect, the backward tunneling
processes is covered in the forward tunneling processes, and the result is
the same as that in \cite{gurvitz}.

Next, the EA effect is taken into account to calculate the
reservoir time correlation functions. It should be noted that the
excitation rate of an electron occupying in lower energy levels
far from the Fermi energy is much smaller than the one near the
Fermi energy. That is, most likely, only the electrons occupying
near the Fermi energy can tunnel through the QPC. If the QPC has a
low transmission, i.e. $n/A,\;m/A\ll \bar{N}$, the states $\left\{
\left| D\left( \bar{L}^{n},R^{n}\right) \right\rangle \right\} $
contributed by all allowed tunnelings in which electrons occupying
near the Fermi energy can tunnel from the source to the drain are
almost the same. Therefore, we have the following approximation
\begin{equation}
\mathrm{Tr}_{D^{(n)}}[\hat{f}_{t}^{+}\hat{f}_{\tau }\hat{\rho}%
_{I}(t)]\approx \hat{\rho}_{f}^{(n)} \sum_{%
\bar{L}^{n}R^{n}} \langle D^{n}|\hat{f}_{t}^{+}\hat{f}%
_{\tau }|D^{n}\rangle ,  \label{eq:assupt1}
\end{equation}
and $C_{f,+}^{n}(t-\tau )$ can be expressed as $\hat{\rho}_{f}^{(n)}(t)\sum_{%
\bar{L}^{n}R^{n}}%
\mathrm{Tr}[\hat{A}_{n}^{+}\hat{f}_{t}^{+}\hat{f}_{\tau }\hat{A}%
_{n}\hat{\rho}_{B}^{(0)}]$, where $\hat{A}_{n}=\hat{a}_{r_{1}}^{+}\cdots
\hat{a}_{r_{n}}^{+}\hat{a}_{\bar{l}_{1}}\cdots \hat{a}_{\bar{l}_{n}}$, and $%
\hat{\rho}_{B}^{(0)}$ is the reservoir vacuum state. Furthermore, we define $%
\hat{\rho}_{B}^{(n)}\equiv \hat{A_{n}}\hat{\rho}_{B}^{(0)}\hat{A_{n}}^{+}$
as the electronic reservoir density operator with $n$ electrons created in
the drain for the forward tunneling processes. Combine these analysis
together, we obtain
\begin{eqnarray}
C_{f,+}^{n}(t-\tau ) &=&\hat{\rho}_{f}^{(n)}(t)\sum_{l^{\prime }r^{\prime
}}\sum_{lr}{e^{\frac{it}{\hbar }\left( \varepsilon _{l^{\prime
}}-\varepsilon _{r^{\prime }}\right) }e^{-\frac{i\tau }{\hbar }\left(
\varepsilon _{l}-\varepsilon _{r}\right) }}  \nonumber \\
&&~~~~~~~\times \sum_{\bar{L}^{n}R^{n}} \mathrm{Tr}\left[ \hat{a}_{r}^{+}\hat{a}_{l}%
\hat{\rho}_{B}^{(n)}\hat{a}_{l^{\prime }}^{+}\hat{a}_{r^{\prime
}}\right]
\nonumber \\
&=&\hat{\rho}_{f}^{(n)}(t)\sum_{lr}e^{\frac{i(t-\tau )}{\hbar }(\varepsilon
_{l}-\varepsilon _{r})}F_{l}^{-}(n,\beta ,t-\tau )  \nonumber \\
&&~~~~~~~~~~~~\times [1-F_{r}^{+}(n,\beta ,t-\tau )],  \label{eq:cfp}
\end{eqnarray}
where the EA-fluctuated Fermi-Dirac functions $F_{l,r}^{\pm }$
have been treated perturbatively and shown
in Eq. (\ref{eq:pf}). Note that Eq.~(\ref{eq:pf}) is valid when $n/A\ll \bar{%
N}$ and $m/A\ll \bar{N}$, which coincides with the approximation used in
Eq.~(\ref{eq:assupt1}). Other reservoir time correlation functions can be
similarly reduced to the following forms
\begin{eqnarray}
&&C_{f,-}^{n}(t-\tau )=\hat{\rho}_{f}^{(n)}(t)\sum_{lr}e^{-\frac{i(t-\tau )}{%
\hbar }(\varepsilon _{l}-\varepsilon _{r})}F_{r}^{+}(n,\beta ,t-\tau )
\nonumber \\
&&~~~~~~~~~~~~~~~~~~~~~~~~~~~~~\times [1-F_{l}^{-}(n,\beta ,t-\tau )],
\nonumber \\
&&C_{b,+}^{m}(t-\tau )=\hat{\rho}_{b}^{(m)}(t)\sum_{lr}e^{\frac{i(t-\tau )}{%
\hbar }(\varepsilon _{l}-\varepsilon _{r})}F_{l}^{+}(n,\beta ,t-\tau )
\nonumber \\
&&~~~~~~~~~~~~~~~~~~~~~~~~~~~~~\times [1-F_{r}^{-}(n,\beta ,t-\tau )],
\nonumber \\
&&C_{b,-}^{m}(t-\tau )=\hat{\rho}_{b}^{(m)}(t)\sum_{lr}e^{-\frac{i(t-\tau )}{%
\hbar }(\varepsilon _{l}-\varepsilon _{r})}F_{r}^{-}(n,\beta ,t-\tau )
\nonumber \\
&&~~~~~~~~~~~~~~~~~~~~~~~~~~~~~\times [1-F_{l}^{+}(n,\beta ,t-\tau )].
\label{eq:cbm}
\end{eqnarray}
Obviously, it shows that, in the above time correlation functions, the
perturbed terms proportional to $\delta \mu _{R,L}(n,\beta ,t)$ come from
the EA effect. Using the perturbation scheme in Sec. II, we then have $\hat{%
\rho}_{f}^{(n)}(t)=\hat{\rho}_{0,f}^{(n)}(t)+\xi \hat{\rho}%
_{1,f}^{(n)}(t)+\xi ^{2}\hat{\rho}_{2,f}^{(n)}(t)+\cdots $ for the forward
tunneling processes, and $\hat{\rho}_{b}^{(m)}(t)=\hat{\rho}%
_{0,b}^{(m)}(t)+\xi \hat{\rho}_{1,b}^{(m)}(t)+\xi ^{2}\hat{\rho}%
_{2,b}^{(m)}(t)+\cdots $ for the backward tunneling processes.

In addition, the time integration in Eq. (\ref{eq:A4}) can be
calculated by using the results
\begin{eqnarray}
&&\frac{1}{2\hbar }{\int_{-\infty }^{\infty }d\tau e^{\frac{\gamma (t-\tau )%
}{i\hbar }}C_{f(b),+}^{n}(t-\tau )}=\pi g_{L}g_{R}\hat{\rho}_{f(b)}^{(n)}(t)
\nonumber \\
&&~~~~~~~~~~~\times \left( \tilde{g}^{(0)}(V_{d}-\gamma )\mp n\mathcal{U}%
\tilde{g}^{(1)}(V_{d}-\gamma )\right) ,  \nonumber \\
&&\frac{1}{2\hbar }{\int_{-\infty }^{\infty }d\tau e^{\frac{i\gamma (t-\tau )%
}{\hbar }}C_{f(b),-}^{n}(t-\tau )}=\pi g_{L}g_{R}\hat{\rho}_{f(b)}^{(n)}(t)
\nonumber \\
&&~~~~~~\times \left( \tilde{g}^{(0)}(-V_{d}+\gamma )\pm n\mathcal{U}\tilde{g%
}^{(1)}(-V_{d}+\gamma )\right) ,  \label{eq:ftm}
\end{eqnarray}
and the relation $\delta \mu _{L}\left( n,\beta ,t\right) +\delta \mu
_{R}\left( n,\beta ,t\right) =n\bar{\mu}\xi \mathcal{F}(t)$, the master
equations become
\begin{widetext}
\begin{eqnarray}
&&\frac{d}{dt}\hat{\rho}_{0,f}^{(n)}(t)+\xi \frac{d}{dt}\hat{\rho}%
_{1,f}^{(n)}(t)=-i\hat{L}_{D}\big(\hat{\rho}_{0,f}^{(n)}(t)+\xi \hat{\rho}%
_{1,f}^{(n)}(t)\big) \,\,\,\,  \nonumber \\
&&~~~~~~-\lambda \Big\{\hat{q}\big(\hat{A}_{+}^{(0)}+\hat{A}_{-}^{(0)}\big)%
\hat{\rho}_{0,f}^{(n)}(t)-\big(\hat{A}_{+}^{(0)}\hat{\rho}_{0,f}^{(n-1)}(t)
+\hat{A}_{-}^{(0)}\hat{\rho}_{0,f}^{(n+1)}(t)\big) \hat{q}  \nonumber \\
&&~~~~~~~~~~~~~~~+\xi \Big[\hat{q}\big(\hat{A}_{+}^{(0)}+\hat{A}_{-}^{(0)}\big)\hat{\rho}%
_{1,f}^{(n)}(t)  -\big(\hat{A}_{+}^{(0)}\hat{\rho}_{1,f}^{(n-1)}(t)+\hat{A}%
_{-}^{(0)}\hat{\rho}_{1,f}^{(n+1)}(t)\big)\hat{q}  \nonumber \\
&&~~~~~~~~~~~~~~~~~~~~~~~~~ +\mathcal{F}(t) n\hat{q}\big(\hat{A}%
_{+}^{(1)}+\hat{A}_{-}^{(1)}\big)\hat{\rho}_{0,f}^{(n)}(t)-\mathcal{F}(t)
\big((n-1) \hat{A}_{+}^{(1)}\hat{\rho}_{0,f}^{(n-1)}(t)
+(n+1)\hat{A}_{-}^{(1)}\hat{\rho}_{0,f}^{(n+1)}(t)\big) \hat{q}
 \Big]\Big\} +H.c. \nonumber \\
&&~~~~~~+\mathcal{O}\left( \xi ^{2}\right) +\cdots \label{eq:m_nf}
\end{eqnarray}
for the forward tunneling processes, and

\begin{eqnarray}
&&\frac{d}{dt}\hat{\rho}_{0,b}^{(n)}(t)+\xi \frac{d}{dt}\hat{\rho}%
_{1,b}^{(n)}(t)=-i\hat{L}_{D}\big(\hat{\rho}_{0,b}^{(n)}(t)+\xi \hat{\rho}%
_{1,b}^{(n)}(t)\big) \,\,\,\,  \nonumber \\
&&~~~~~~-\lambda \Big\{\hat{q}\big(\hat{A}_{+}^{(0)}+\hat{A}_{-}^{(0)}\big)%
\hat{\rho}_{0,b}^{(n)}(t)-\big(\hat{A}_{-}^{(0)}\hat{\rho}_{0,b}^{(n-1)}(t)
+\hat{A}_{+}^{(0)}\hat{\rho}_{0,b}^{(n+1)}(t)\big) \hat{q}  \nonumber \\
&&~~~~~~~~~~~~~~~+\xi %
\Big[\hat{q}\big(\hat{A}_{+}^{(0)}+\hat{A}_{-}^{(0)}\big)\hat{\rho}%
_{1,b}^{(n)}(t)-\big(\hat{A}_{-}^{(0)}\hat{\rho}_{1,b}^{(n-1)}(t)+\hat{A}%
_{+}^{(0)}\hat{\rho}_{1,b}^{\left( n+1\right) }(t)\big)\hat{q}  \nonumber \\
&&~~~~~~~~~~~~~~~~~~~~~~~~~ -\mathcal{F}(t) n\hat{q}\big(%
\hat{A}_{+}^{(1)}+\hat{A}_{-}^{(1)}\big)\hat{\rho}_{0,b}^{(n)}(t)+\mathcal{F}(t) \big((n-1)%
\hat{A}_{-}^{(1)}\hat{\rho}_{0,b}^{(n-1)}(t)+(n+1)\hat{A}_{+}^{(1)}
\hat{\rho}_{0,b}^{(n+1)}(t)\big) \hat{q}\Big]\Big\}+H.c. \nonumber \\
&&~~~~~~+\mathcal{O}\left( \xi ^{2}\right) +\cdots \label{eq:m_nb}
\end{eqnarray}
\end{widetext}
for the backward tunneling processes, where $\hat{A}_{\pm }^{\left(
0,1\right) }=\sum_{i=0,1,2}G_{\pm ,i}^{\left( 0,1\right) }\hat{P}_{i}$. All
the definitions of the elements of Eqs.~(\ref{eq:m_nf},\ref{eq:m_nb}) can be
found in Sec. II.

Eqs.~(\ref{eq:m_nf},\ref{eq:m_nb}) describes the transport properties of
electrons in the system we concerned, for example, the transport current and
noise spectrum \cite{gurvitz,korotkov,goan,mozyrsky,stace,li}. According to
Eq.~(\ref{eq:tstat}) and recalling that $\hat{\rho}_{f(b)}^{(n)}(t)$
describes the quantum oscillation of the electron in the CQD with the
conditions of $n$ electrons accumulating in the drain (source), the $k$-th
order perturbation of the qubit oscillation is completely described by $\hat{%
\rho}_{k}(t)={\sum_{n=0}^{\infty }}\hat{\rho}_{k,f}^{(n)}(t)+{%
\sum_{m=0}^{\infty }}\hat{\rho}_{k,b}^{(m)}(t).$ Accordingly, the $k$-th
order perturbation of the QPC current operator is given by $\frac{d\hat{N}%
_{k}}{dt},$ where $\hat{N}_{k}={\sum_{n=0}^{\infty }}n\hat{\rho}%
_{k,f}^{(n)}(t)-{\sum_{m=0}^{\infty }}m\hat{\rho}_{k,b}^{(m)}(t),$ and the $k
$-th order perturbation of noise spectrum is defined by $\hat{W}_{k}(t)={%
\sum_{n=0}^{\infty }}n^{2}\left( \hat{\rho}_{k,f}^{(n)}(t)+\hat{\rho}%
_{k,b}^{(k)}(t)\right) $. Combine with Eqs. (\ref{eq:m_nf},\ref{eq:m_nb}),
one can obtain the master equations shown in Sec. II straightforwardly.

\acknowledgements

One of the authors (M.T.L) would like to thank Profs. B. L. Hu and C. E. Lee
for useful discussions. The work is supported by the National Science
Council of Republic of China under Contract Nos. NSC-93-2119-M-006-002 and
NSC-93-2120-M-006-005, and National Center for Theoretical Science, Republic
of China.

\end{document}